\documentclass[structabstract]{aa}  
\usepackage{graphicx}
\usepackage{txfonts}
\begin{document}
   \title{The galaxy stellar mass function at $3.5\le z\le 7.5$ in the
CANDELS/UDS, GOODS-South, and HUDF fields}

\author{A. Grazian \inst{1}
\and
A. Fontana \inst{1}
\and
P. Santini \inst{1}
\and
J. S. Dunlop \inst{2}
\and
H. C. Ferguson \inst{3}
\and
M. Castellano \inst{1}
\and
R. Amorin \inst{1}
\and
M. L. N. Ashby \inst{4}
\and
G. Barro \inst{5}
\and
P. Behroozi \inst{3}
\and
K. Boutsia \inst{1}
\and
K. I. Caputi \inst{6}
\and
R. R. Chary \inst{7}
\and
A. Dekel \inst{8}
\and
M. A. Dickinson \inst{9}
\and
S. M. Faber \inst{5}
\and
G. G. Fazio \inst{4}
\and
S. L. Finkelstein \inst{10}
\and
A. Galametz \inst{11}
\and
E. Giallongo \inst{1}
\and
M. Giavalisco \inst{12}
\and
N. A. Grogin \inst{3}
\and
Y. Guo \inst{5}
\and
D. Kocevski \inst{13}
\and
A. M. Koekemoer \inst{3}
\and
D. C. Koo \inst{5}
\and
K.-S. Lee \inst{14}
\and
Y. Lu \inst{15}
\and
E. Merlin \inst{1}
\and
B. Mobasher \inst{16}
\and
M. Nonino \inst{17}
\and
C. Papovich \inst{18}
\and
D. Paris \inst{1}
\and
L. Pentericci \inst{1}
\and
N. Reddy \inst{16}
\and
A. Renzini \inst{19}
\and
B. Salmon \inst{18}
\and
M. Salvato \inst{11}
\and
V. Sommariva \inst{20}
\and
M. Song \inst{10}
\and
E. Vanzella \inst{21}
          }

   \offprints{A. Grazian, \email{andrea.grazian@oa-roma.inaf.it}}

\institute{INAF - Osservatorio Astronomico di Roma, Via Frascati 33,
I--00040, Monteporzio, Italy
\and SUPA, Institute for Astronomy, University of Edinburgh,
Royal Observatory, Edinburgh EH9 3HJ, UK
\and Space Telescope Science Institute, 3700 San Martin Drive, Baltimore,
MD 21218, USA
\and Harvard-Smithsonian Center for Astrophysics
60 Garden Street, Cambridge, MA 02138, USA
\and UCO/Lick Observatory, University of California, 1156 High
Street, Santa Cruz, CA 95064, USA
\and Kapteyn Astronomical Institute, University of Groningen,
9700 AV Groningen, The Netherlands
\and California Institute of Technology, Pasadena, CA 91125, USA
\and Center for Astrophysics and Planetary Science, Racah Institute of
Physics, The Hebrew University, Jerusalem 91904, Israel
\and NOAO, 950 N. Cherry Avenue, Tucson, AZ 85719, USA
\and Department of Astronomy, The University of Texas at Austin, Austin,
TX 78712, USA
\and Max Planck Institute for extraterrestrial Physics, Giessenbachstrasse 1,
D-85748 Garching bei Munchen, Germany
\and Department of Astronomy, University of Massachusetts, 710 North
Pleasant Street, Amherst, MA 01003, USA
\and Department of Physics and Astronomy, University of Kentucky,
Lexington, KY 40506, USA
\and Physics Department Purdue University,
525 Northwestern Avenue West Lafayette, IN 47907, USA
\and Kavli Institute for Particle Astrophysics \& Cosmology, Physics
Department, and SLAC National Accelerator Laboratory, Stanford University,
Stanford, CA 94305, USA
\and Department of Physics and Astronomy, UC Riverside, 900 University Ave,
Riverside, CA 92521, USA
\and INAF - Osservatorio Astronomico di Trieste, Via G.B. Tiepolo 11,
I--34131, Trieste, Italy
\and George P. and Cynthia Woods Mitchell Institute for Fundamental Physics and
Astronomy, and Department of Physics and Astronomy, Texas A\&M University,
College Station, TX 77843-4242, USA
\and INAF - Osservatorio Astronomico di Padova, vicolo dell’Osservatorio 5,
I-35122 Padova, Italy
\and University of Bologna, Department of Physics and Astronomy (DIFA), V.le
Berti Pichat 6/2, I-40127, Bologna, Italy
\and INAF - Osservatorio Astronomico di Bologna, via Ranzani 1, I-40127
Bologna, Italy
}

   \date{Received Month day, year; accepted Month day, year}

   \authorrunning{Grazian et al.}
   \titlerunning{The high-z stellar mass function in CANDELS}

  \abstract
{The form and evolution of the galaxy stellar mass function at high
redshifts provide key information on star-formation history and mass
assembly in the young Universe, close or even prior to the epoch of
reionization.}
{We aimed to use the unique combination of deep
optical/near-infrared/mid-infrared imaging provided by HST,
Spitzer and the VLT in the CANDELS-UDS, GOODS-South, and HUDF fields
to determine the galaxy stellar mass function (GSMF) over
the redshift range $3.5\le z\le 7.5$.}
{We utilised the HST WFC3/IR near-infrared imaging from CANDELS
and HUDF09, reaching $H\simeq 27-28.5$ over a total area of 369
arcmin$^2$, in combination with associated deep HST ACS optical
data, deep Spitzer IRAC imaging from the SEDS programme, and
deep $Y$ and $K$-band VLT Hawk-I images from the HUGS programme, to
select a galaxy sample with high-quality photometric redshifts. These
have been calibrated with more than 150 spectroscopic redshifts in the
range $3.5\le z\le 7.5$,
resulting in an overall precision of $\sigma_z/(1+z)\sim 0.037$. With
this database we have determined the low-mass end of the high-redshift
GSMF with unprecedented precision, reaching down to masses as low as
$M^{*}\sim 10^9\,{\rm M_{\odot}}$ at $z=4$ and $\sim 6\times 10^9\,
{\rm M_{\odot}}$ at $z=7$.}
{We find that the GSMF at $3.5\le z\le 7.5$ depends only slightly on
the recipes adopted to measure the stellar masses, namely the
photometric redshifts, the star-formation histories, the nebular
contribution or the presence of AGN on the parent sample. The low-mass
end of the GSMF is steeper than has been found at lower redshifts, but
appears to be unchanged over the redshift range probed here. Meanwhile
the high-mass end of the GSMF appears to evolve primarily in density,
albeit there is also some evidence for evolution in characteristic
mass. Our results are very different from previous mass function
estimates based on converting UV galaxy luminosity functions into mass
functions via tight mass-to-light relations. Integrating our evolving
GSMF over mass, we find that the growth of stellar mass density is
barely consistent with the time-integral of the star-formation rate
density over cosmic time at $z>4$.}
{These results confirm the unique synergy of the CANDELS+HUDF, HUGS,
and SEDS surveys for the discovery and study of moderate/low-mass
galaxies at high redshifts, and reaffirm the importance of space-based
infrared selection for the unbiased measurement of the evolving galaxy
stellar mass function in the young Universe.}


   \maketitle
%

\section{Introduction}

Deep multi-wavelength surveys have rapidly 
expanded our knowledge of the young Universe, 
with the most recent deep near-infrared imaging 
pushing back the redshift frontier of photometrically-selected 
galaxies out to $z\simeq 7-12$ (\cite{dunlop13,coez11,ellisz12,oeschz12}).
Moreover, the physical properties of high-redshift galaxies, such 
as their star-formation rates (SFR) and stellar masses,  
can now be determined with meaningful accuracy up to 
$z\simeq 7-8$ (\cite{labbe10,labbe13}), thanks to the combined power of 
deep space-based ({\it HST}, {\it Spitzer}) and ground-based (VLT) 
near-mid infrared imaging (\cite{goodsK,hugs,seds}).

The evolution of star-formation activity in galaxies over cosmic 
history, and the physical processes which may drive and limit such activity, 
have been the subject of intensive observational and theoretical study 
in recent years (for a review see \cite{md14}).
The ultimate goal of the latest generation of galaxy-formation models
is to represent, with fully-developed cosmological simulations in the
$\Lambda$-CDM framework, the baryonic assembly of structures at
different mass scales in the Universe as a function of cosmic time.
While dark matter evolution is rather simple and clear, the physics
regulating the baryonic processes is complex to model, and non-trivial 
to understand: star formation mechanisms, gaseous dissipation,
feedback from stars and active galactic nuclei (AGN), turbulence, 
and the role of mergers, are only some of the many problems 
encountered when trying to build a fully-realistic simulation of 
galaxy formation and evolution (\cite{springel10}).

Stellar mass is a physical parameter that provides a useful and
complementary view of galaxy evolution from the measurement of
SFR. From an observational perspective, given infrared data of
sufficient quality and depth, stellar mass is a more straightforward
and robust quantity to measure, being less subject to degenerate
uncertainties in age, metallicity, and dust extinction.  From the
theoretical point of view, since stellar mass is a time-integrated
quantity, it is less sensitive to the details of the star formation
history (i.e. bursts of star formation, SF quenching). The detailed
measurement of the growth of the stellar mass content in galaxies thus
offers a key observational probe of the underlying physical processes
driving and limiting star-formation activity throughout cosmic
time. Two statistical descriptions are often used to quantify the
growth/distribution of stellar mass as a function of redshift: the
Galaxy Stellar Mass Function (GSMF) and its integral over mass, namely
the Stellar Mass Density (SMD hereafter).

Recently, ground-based optical and NIR surveys (e.g. SDSS, UKIDSS,
UltraVISTA), have been successfully utilised to explore the physical
properties of galaxies at low redshift and to extend such studies out 
to high-redshift for the high-mass/high-luminosity tail 
of the galaxy distribution. For example, in the local Universe,
\cite{baldry2012} and \cite{moustakas2013} have studied the
GSMF down to $M\sim 10^8 M_\odot$, while
\cite{ilbert13} and \cite{muzzin13} have extended the GSMF studies to 
$z \simeq 4$, albeit inevitably limited to progressively 
higher stellar masses ($M>10^{10}\,{\rm M_\odot}$).

While the recent ground-based progress is impressive, it remains the
case that the analysis of the low-mass tail of the GSMF, especially at
high redshifts, requires very deep infrared imaging, which is only
really possible with space-based instrumentation (i.e. {\it HST} and
{\it Spitzer}). Accordingly, several studies have used deep {\it HST}
and {\it Spitzer} data to begin to investigate the GSMF at $z\ge 4$
(\cite{stark09,labbe10,gonzalez2011,caputi11,lee2012,duncan14}) or
down to very small masses but at $z\le 4$ (\cite{tomczak13}).

Despite this important progress, the robust study of the
GSMF at early cosmological epochs has been seriously hampered 
by a lack of appropriately deep near--mid-infrared imaging 
over sufficiently large areas of sky. Consequently, there remain 
many important outstanding questions to be resolved concerning the 
assembly history of Universe, as quantified through the form and evolution
of the the GSMF. Examples of current key issues, over which 
there remains considerable controversy and confusion, are given below:
 
\begin{itemize}
\item
Is the high-mass end of the GSMF evolving at $0<z<3$?  Both
\cite{pg08} and \cite{ilbert13} found that the GSMF at
$M>10^{11.7}\,{\rm M_{\odot}}$ did not evolve strongly from $z\simeq
2$ to $z\simeq 0.3$, while \cite{marchesini09} showed that the SMD of
these galaxies evolved by a factor of $\sim 50$ in the last 10\,Gyr.
\item
Does the low-mass end of the GSMF steepen at high-redshift? Recent attempts
to extend the study of the GSMF
to high redshifts and low masses have produced contrasting results.
For example, \cite{santini12} found a steepening towards high-redshift, while
\cite{ilbert13} found no evidence that the low-mass end 
of the GSMF was evolving in shape. Uncertainty over the faint-end
slope of the GSMF is key to the next open problem,
the measurement of the growth of SMD.
\item
Is the integral of the SFRD consistent with the observed SMD ?  The
time integral of the SFRD, corrected for gas recycling fraction (i.e.
gas lost by aging stars), has been claimed to exceed the measured SMD
(\cite{wilkins08,rs09,santini12}). As shown by \cite{rs09}, this
apparent conflict could possibly be resolved by properly matching the
integration limits in the UV galaxy luminosity function (LF) and in
the GSMF. Subsequently, \cite{santini12} proposed that a steepening of
the GSMF with increasing redshift might remove any discrepancy at
$2<z<4$, but at $z<2$ found that the steepening was insufficient to
bridge the apparent gap. More recently still, \cite{behroozi13}
pointed out that the previous estimates of SFR density over cosmic
time by \cite{hb06} may have been overestimated, due to potentially
excessive corrections for dust extinction when inferring SFR, but as
discussed by Madau \& Dickinson (2014), some tension still remains at $z<2$.
\item
Do colour and photo-z galaxy selection methods provide a consistent
sampling of the GSMF ? \cite{rs09} suggested that up to $\simeq 50\%$
of the total stellar mass in the redshift range $1.9<z<3.4$ is in
faint galaxies with stellar masses smaller than $\sim 10^{10}\,{\rm
M_{\sun}}$, as compared to $\simeq 10-20\%$ as obtained from an
extrapolation of the Schechter fit to the observed MF obtained by
\cite{marchesini09}. It is worth noticing that \cite{rs09} converted
UV luminosity directly into stellar mass. At higher redshifts,
\cite{gonzalez2011} converted the observed UV luminosity function of
Lyman-break galaxies at $z>3$ into a GSMF using similar assumptions on
the mass-to-light ratio.  Using a somewhat different approach,
\cite{lee2012} derived the mass function of UV-selected LBGs at $z\sim
4-5$, finding a flatter ($\alpha\sim -1.3$) slope with respect to the
UV luminosity function of star-forming galaxies at the same redshifts.
Their results do not fully agree with the GSMFs derived from NIR or
MIR selected samples (\cite{pg08,marchesini09,marchesini10,caputi11,
santini12}). Recently, \cite{duncan14} addressed the issue of
photo-z selection vs Lyman-break selection, showing that the two
methods are almost equivalent, once the photometric scatter is
properly treated.
\item
Which are the most appropriate stellar libraries to use when computing
the stellar mass in galaxies, especially at high-redshift?
\cite{maraston05,cb07} (M05, BC07 hereafter) showed that the TP-AGB
phase could have a strong impact on the stellar mass derivation from
infrared light, especially in the redshift range $0.5<z<2.0$. This
problem has also been investigated by \cite{henriques11} and recently
this topic has been the subject of a number of papers
(\cite{tonini10,massreview,mobasher14}), indicating that this is an
important issue in the GSMF field.
\item
What is the impact of AGNs on the high-mass end of the GSMF?
\cite{fontana06} excluded all
AGNs from their sample, while \cite{marchesini09} included AGNs in their
sample, obtaining slightly different results, especially at the massive
tail of the GSMF. \cite{santini12} included only type 2 AGNs in the ERS field.
As shown by \cite{santini12b}, both for type-1 and type-2
AGN at $z\le 2.5$, the stellar mass derived by adopting only stellar libraries
showed no systematic offset from the one coming from a two-component
fit (stars+AGN), but presented a large spread (RMS of 0.34 dex for type 1
AGN). This has been explained by the fact that the AGN
(especially type 1) is providing additional non-stellar light but it
is also making the SED bluer than the pure stellar one. In this case
the additional light by the AGN is compensated by the lower M/L ratio.
\item
What are the contributions of the nebular lines and continuum to the SED
of high-redshift galaxies and hence on the derived stellar masses?
Recently, several studies have endeavored to include 
the contribution of nebular lines and continuum in the fitting 
of high-redshift galaxy SEDs. In
particular, \cite{schaerer2009} showed that the model fit to 
the SEDs of $z>3$ galaxies can be significantly improved by the
inclusion of the nebular lines and continuum, and inclusion of 
this contribution also helps to yield more reasonable ages for galaxies 
at very high redshifts. More recently, the importance of including 
the nebular contribution has been inferred more directly from 
observations via analysis of the Spitzer-IRAC photometry of 
high-redshift galaxies. \cite{shim2011} reported
a strong H$\alpha$ line contribution to the Spitzer IRAC 3.6$\mu$m and 4.5$\mu$m
bands for a small sample of galaxies with spectroscopic
redshifts at $z\simeq 4$. Further studies (e.g. 
\cite{stark13,labbe13,oesch13,schenker13}) have shown 
that the nebular contribution can also be important at $z>3-4$, when the 
[OIII] and H$\beta$ lines enter the IRAC 3.6$\mu$m filter.
Thus, the inclusion of the nebular contribution (both lines and continuum)
is becoming progressively more common in the SED fitting of the photometry 
of high-redshift galaxies (e.g. \cite{salmon13}), 
although how best to estimate the appropriate level 
of nebular contribution remains a matter of debate.
\item
Are theoretical models able to reproduce the observed GSMF?
A common feature of predictions from $\Lambda$-CDM models has been an 
over-production of low-mass galaxies, especially at high redshifts
(\cite{wang2008,bielby2012,bower2012,guo2011}).
Recently, \cite{lu13} claimed a better agreement of the recent
renditions of SAMs with the observed GSMF at all redshifts ($z=0-6$).
However, as pointed out by \cite{ilbert13}, the theoretical predictions 
are still far from reproducing the GSMF of the old/evolved population
and the disagreement is larger for the higher redshifts.
Using simulations, \cite{wilkins13} recently predicted the properties of
high-redshift galaxies, but were unable to fully reproduce the observed GSMF
of \cite{gonzalez2011}, both for the low-mass galaxies at $z \simeq 5$ and
for the higher-mass galaxies at $z \simeq 7$.
\end{itemize}

The CANDELS project (\cite{koekemoer11,grogin11}), with its particular
combination of survey volume, depth and wavelength coverage (0.5-1
$\mu$m rest-frame), provides an ideal dataset with which to attempt to
resolve some of these issues. In this paper we use the CANDELS data to
investigate a number of the outstanding issues mentioned above,
exploring carefully how stellar mass derivation depends on the recipes
used to derive photometric redshifts, the assumed galaxy star
formation histories, the nebular contribution, the AGN content of
galaxy samples, and field-to-field variations in the galaxy samples.
We then derive and present a new robust analysis of the form and
evolution of the GSMF at high-redshift ($z\ge 3.5$).

This paper is organized as follows: after introducing the photometric
and spectroscopic dataset in Section 2, we present the stellar mass
estimates in Section 3.1. The derivation of the stellar mass function
is discussed in detail in Section 3.2, and Section 4 is devoted to
determining the uncertainties on the GSMF estimate. We present our
results in Section 5, and include an analysis of the shape of the
GSMF, a comparison with recent results in the literature, a discussion
of the mass-to-light ratio of galaxies at $z\simeq 4$ and an
investigation of the inferred physical properties of massive galaxies
at high redshift. Section 6 describes the redshift evolution of the
GSMF, while the stellar mass density and its comparison with the
integrated SFRD is discussed in Section 7. Finally, we summarize our
results in Section 8. In Appendix A we compare different recipes for
the calculation of the GSMF, in Appendix B we describe the correction
of the Eddington bias, and in Appendix C for completeness we provide
the results obtained by neglecting the Eddington bias correction.
Throughout we adopt the $\Lambda$-CDM concordance cosmological model
($H_0 = 70 {\rm km s^{-1} Mpc^{-1}}$, $\Omega_M=0.3$ and
$\Omega_\Lambda=0.7$). All magnitudes are in the AB system, and a
\cite{salpeter} stellar initial mass function (IMF) is assumed in the
derivation of all galaxy masses.


\section{Data}

\subsection{The photometric dataset}

The CANDELS survey (\cite{grogin11,koekemoer11}) is an ideal dataset
with which to study the stellar masses of high-redshift galaxies,
thanks to its combination of deep photometry and reasonably wide areal
coverage, with superb image quality obtained with the near-infrared
camera on {\it HST}, the Wide Field Camera 3
(WFC3\footnote{http://www.stsci.edu/hst/wfc3}).  We have used the
first two CANDELS fields, namely the CANDELS-Wide imaging within the
UKIDSS Ultra Deep Survey (UDS) (covering $\simeq 200\,{\rm arcmin^2}$
to a 5-$\sigma$ depth of $H_{160}=26.7$) and the maximum depth/area CANDELS
imaging of the GOODS-South field (covering $\simeq 170\,{\rm
arcmin^2}$ to a mean 5-$\sigma$ depth of $H_{160}=27.5$), combining data
from the ERS, CANDELS-Wide, CANDELS-Deep, and the main pointing of the
HUDF09 program ($\simeq 5\,{\rm arcmin^2}$ down to 5-$\sigma$
$H_{160}=28.5$). This data set does not include the two parallel fields of
the HUDF09 program, nor the HUDF12 data (\cite{koekemoer13}). The
CANDELS-UDS and GOODS-South+HUDF09 fields with their associated
multi-wavelength catalogues are fully described in \cite{galametz} and
\cite{guo}, respectively.

These imaging data, of unprecedented quality and depth, provide a
powerful dataset for stellar mass function investigations, especially
at low masses and at high redshifts. In particular, they include very
deep imaging with the IRAC instrument aboard the {\sl Spitzer Space
Telescope} from the {\sl Spitzer Extended Deep Survey} (SEDS;
\cite{seds}), covering the CANDELS fields to a 3-$\sigma$ depth of
26\,AB mag at both 3.6 and 4.5\,$\mu$m, that are crucial for sampling
the rest--frame optical bands at $z>4$.

Another crucial data set that is unique to these two fields is the
deep Hawk-I imaging obtained through the HUGS (Hawk-I UDS and GOODS
Survey) VLT programme (\cite{hugs}). This has delivered deep
ground-based $Y$ and $K$-band images of a depth well matched to the
$H$-band magnitude limits of the CANDELS survey, with exposure times
ranging from $\simeq 12$ hours over the shallower CANDELS images to
about 85 hours of integration in the deepest region of the
HUGS/GOODS-South field (which includes most of the Hubble Ultra Deep
Field). In the deepest area of GOODS-South, the HUGS data reach a
1$\sigma$ magnitude limit per square arcsec of $\simeq$28.0~mag in the
$K$ band. The $Y$ and $K$-band imaging of the UDS field reaches a
1$\sigma$ magnitude limit per square arcsec of $\simeq$28.3 and 27.3,
respectively. The image quality of the HUGS images is extremely good,
with a seeing of $0.37-0.43$\,arcsec in the $K$ band, and only
slightly poorer in the $Y$ band ($0.45-0.50$\,arcsec). This makes the
HUGS survey the deepest $K$-band image over a significant area
($>340\,{\rm arcmin^2}$), the only deeper $K$-band imaging being the
Super Subaru Deep Field (\cite{ssdf}), which covers only a very small
area ($\simeq 1\,{\rm arcmin^2}$) with the aid of adaptive optics. In
\cite{hugs} we show that in the HUGS $K$-band data we can detect, at
1$\sigma$, more than 90\% of the $H_{160}$-band detected galaxies in
CANDELS.

The final HUGS data have already been included in the official CANDELS
catalogue produced by \cite{galametz}, but in the GOODS-South
catalogue produced by \cite{guo}, only a fraction of the deep $K$-band
imaging from HUGS was included. In this work we use instead the final
version of the HUGS $K$-band image in the GOODS-South field, analysed
with the same TFIT code for deep blended photometry. This resulting
catalogue is fully described in \cite{hugs}.

In addition, we further enhanced the official CANDELS GOODS-South
catalogue presented by \cite{guo} by adding the deep VIMOS $B$-band
imaging in the field (\cite{sommariva14,nonino}). This imaging has an
effective wavelength of $4310$ \AA, slightly bluer than the B435W ACS
filter on-board {\it HST} (4350 \AA). The mean seeing of the
ground-based image is 0.8\,arcsec, but the combination of exposure
time (28 hours) and collecting area of the 8.2m VLT telescope yields a
magnitude limit of 30.5 mag, (1-$\sigma$) which is much deeper than
that reached by the HST-ACS B435 imaging (\cite{giavalisco04}), which
is 29.2 mag (1-$\sigma$). The $B$-band VIMOS photometry has been
computed with the TFIT software with the same technique adopted for
the other ground-based bands, as described by \cite{guo}.

Both the deep $B$ and $K$-band ground-based images in GOODS-South have
been included with the aim of better constraining the galaxy SEDs and
the stellar masses, but they have not been used for refining the
photometric redshift solutions.  For the latter, we adopt the results
presented by \cite{dahlen13}, as described further below.

\subsection{Photometric catalogue}

We have used the official CANDELS catalogues in the GOODS-South and UDS fields,
where object selection has been performed in the $H_{160}$ band of
WFC3. The total number of sources detected in the CANDELS UDS and
GOODS-South fields are 35932 and 34930, respectively.

Due to the complexity of the GOODS-South and UDS exposure maps in the
$H_{160}$ band, the magnitude limit varies over the field, such
that it is impossible to assign a single completeness limit to the
whole survey. To overcome this limitation we divided the survey
in five areas with relatively homogeneous magnitude limits. To
achieve this, we converted the absolute RMS maps associated with the
$H_{160}$ science frames into magnitude limit maps at 1-$\sigma$ and in a
given area of the sky (1 arcsec$^2$). This value is used as a
conventional reference limit to define the various regions of
different depths for each galaxy in the two fields. For the UDS,
details can be found in Fig.3 of \cite{galametz}, while for
GOODS-South the reference plot is Fig.1 of \cite{guo}.

To associate a proper completeness magnitude to each area of the
survey (defined by a range of magnitude limits at 1-$\sigma$ in
an area of 1 arcsec$^2$), we ran simulations adding artificial
sources (point-like) to the $H_{160}$ band images and recovered them
using the same SExtractor configuration adopted in \cite{galametz} and
\cite{guo}. Then, we computed the completeness in magnitude and
derived, at a given flux limit (at 1-$\sigma$ in an area of 1
arcsec$^2$), the magnitude at which the completeness is above
90\%. This allows us to simply associate to each galaxy, given
its magnitude limit computed locally, a proper completeness limit.

We consider here primarily the detection completeness, although there
are also systematic effects on the magnitude estimate, especially
for fainter objects. If we restrict the analysis to sources which are
0.5 magnitude brighter than the completeness limit, we find that 90\%
of them are within 0.2 magnitudes of the input flux value in
our simulations. This fraction then goes to $\simeq 75\%$ approaching
the completeness limit. We have verified that this limit is comparable
with the independent results of \cite{duncan14} for the GOODS-South
field. In the following, we will restrict our analysis to brighter than 
this value, hereafter called photometric limit, to distinguish it from the mass
completeness limit that will be discussed later. In any case the cut
that will be applied to ensure a robust completeness in mass, as we
discuss later, is brighter than the photometric completeness limit due to the
detection of objects, described here. Table \ref{table:magcompl}
provides the depth and the area covered by different sub-regions. It
also summarizes the number of galaxies available in each region
and the total number, which is $\simeq$55,000.

\begin{table*}
\caption{Area and magnitude limits of the CANDELS GOODS-South, HUDF,
and UDS fields}
\label{table:magcompl}
\centering
\begin{tabular}{c | r c c r r}
\hline\hline
Field & Area & $H_{160}$ Mag. limit & $H_{160}$ Compl. limit & $N_{gal}$ &
$N^{high-z}_{gal}$ \\
 & arcmin$^2$ & 1$\sigma$ in 1 arcsec$^2$ & 90\% & & $3.5<z<7.5$ \\
\hline
GOODS-South \#1 &  11.05 & $27.00<H_{160}<28.08$ & 26.00 &   801 &   44 \\
GOODS-South \#2 &  25.03 & $28.08<H_{160}<28.32$ & 26.25 &  2794 &  132 \\
GOODS-South \#3 &  50.47 & $28.32<H_{160}<28.83$ & 26.75 &  7984 &  495 \\
GOODS-South \#4 &  77.18 & $28.83<H_{160}<29.40$ & 27.25 & 15310 & 1231 \\
GOODS-South \#5 (HUDF) &   5.18 & $29.40<H_{160}<31.00$ & 28.00 &  1672 &  132 \\
\hline
UDS \#1 &  58.02 & $26.00<H_{160}<27.90$ & 26.10 &  5734 &  311 \\
UDS \#2 & 131.70 & $27.90<H_{160}<28.20$ & 26.40 & 18986 &  903 \\
UDS \#3 &  10.27 & $28.20<H_{160}<30.00$ & 26.70 &  1358 &   59 \\
\hline
TOTAL   & 368.90 & - & - & 54639 & 3307 \\
\hline
\end{tabular}
\end{table*}

A more general treatment of the completeness would require repetition of the
sample selection, simulating all the photometric bands, and redoing the
analysis with TFIT on the low-resolution images, verifying the effects
on the photometric redshifts and mass estimation. This is however
beyond the scope of this paper, and we refer to \cite{lee2012} for a
demonstration of the robustness of our photometric approach.

\subsection{Spectroscopic and photometric redshifts}

The two catalogues (GOODS-South and UDS) were cross-correlated to
existing spectroscopic samples, as described in \cite{galametz} and
\cite{guo}. Additional spectroscopic redshifts were added to the
present sample by a collection of high-redshift LBGs from
\cite{fontana10,vanzella11,pentericci11} and preliminary results of
the ESO Large Programme (PI L. Pentericci) with 140 expected hours of
FORS2 spectroscopy on 3 CANDELS fields (UDS, COSMOS,
GOODS-South). Currently, we have 31 $z>5.5$ spectroscopic redshifts in
the GOODS-South field, a number which is 3 times larger than the
public spectroscopic redshifts currently available on this field (9
galaxies only). In total, there are 2272 spectroscopic redshifts of
good quality in the GOODS-South field, and 308 in the CANDELS-UDS in
the $0<z<7$ interval. Restricting the sample to $3.5<z<7.5$, there
are in total 152 galaxies with robust spectroscopic redshifts. As
shown later in Sect.3, these objects sample the redshift-mass plane
with reasonable completeness up to $z\simeq 6$, and become rare at $z>6$.

For sources lacking spectroscopic information, photometric redshifts
were computed by optimally-combining six different photometric
redshifts, as described in \cite{dahlen13}. All these photometric
redshifts were computed by fitting the observed spectral energy
distribution (SED) of the objects from the $U$ band to the 8.0\,$\mu$m
band of {\it Spitzer} using different codes and synthetic
libraries. Using a training sample of 1193 spectroscopic redshifts,
the optimal photometric redshift solution has been derived by taking
into account small zero-point offsets and adding extra smoothing
errors to the individual probability distribution function (PDF) in
redshift. Then a unique PDF as a function of redshift was derived by
optimally combining the 6 individual PDFs using a hierarchical
Bayesian approach as explained in \cite{dahlen13}.  This method
significantly improves the final accuracy compared to the individual
recipes. The absolute scatter of $|\Delta z|/(1+zspec)$ is equal to
0.03, with only 3.4\% of ``catastrophic'' outliers (defined as objects
with $|\Delta z|/(1+zspec)> 0.15$), when the comparison is made with
the spectroscopic training set, at relatively bright magnitudes
($H_{160}\le 24$). Comparing the spectroscopic redshifts at
$3.5<z<7.5$ with the Bayesian photometric redshifts we find a scatter
of $\sigma_z/(1+z)=0.037$ and an outlier fraction of 11 out of 152
objects (7.2\%). The tests with the galaxy pairs show that the
uncertainty increases to about 0.06 at $H_{160}\simeq 26$
(\cite{dahlen13}). This method also delivers the redshift probability
distribution PDF(z) for each galaxy, that is then used to estimate the
relevant uncertainties in the stellar mass functions, as explained in
the following sections.

In total, there are 2034 galaxies in the GOODS-South field and 1273
galaxies in UDS with a robust spectroscopic redshift or, alternatively,
with a photometric redshift in
the range $3.5<z<7.5$. This defines the sample which will be analysed
in this paper with the aim of deriving the GSMF over this redshift
range. We note that this sample represents a small sub-sample of the
total CANDELS GOODS-South and UDS galaxy catalogues (7\% and 5\%
respectively).


\section{The derivation of the Galaxy Stellar Mass Function in the CANDELS
fields}

\subsection{Stellar masses}
\label{sec:stellarmasses}

We have derived the stellar masses using a spectral-fitting
technique similar to that used in previous studies
(\cite{fontana04,grazian06a,fontana06,amaze} and \cite{santini12}),
and similar to those adopted by other groups in the literature
(e.g. \cite{dickinson03}, \cite{drory04},
\cite{ilbert13,muzzin13,tomczak13}).

More precisely, to derive the stellar mass of each galaxy, we have
fitted the observed SED after fixing the redshift to the high-quality
spectroscopic value, or to the photometric one when the former is not
available or is not robust. The SED fitting method is based on the
$\chi^2$ minimization of the differences between the observed
multicolour distribution of each object and a set of templates,
computed with standard spectral synthesis models (\cite{bc03} in our
case, hereafter BC03). The adopted synthetic library broadly
encompasses the variety of star-formation histories, metallicities
and extinctions displayed by real galaxies. To facilitate the comparison with
previous studies, we have used the Salpeter IMF, ranging over a set of
metallicities (from $Z=0.02 Z_\odot$ to $Z=2.5 Z_\odot$) and dust
extinction ($0<E(B-V)<1.1$), with a \cite{Calzetti2000} or a Small
Magellanic Cloud (\cite{smc}) extinction curve left as a free
parameter. Different star-formation histories (SFH) have been
adopted, as described below. In all cases the age is defined as the
time elapsed since the onset of star formation, and at each redshift
this is varied within a fine grid, the only constraint being that it must be
lower than the age of the Universe at that redshift. As in previous analyses,
the derived stellar masses are corrected for the gas recycling
fraction, (i.e. the fraction of baryons that are returned to the ISM
because of stellar winds and SN explosions) taking into account the
recipes of BC03. We thus do not use the total integral of gas turned
into stars, but only the mass which is actually in the form of stars.
For each model in the adopted library, we have
computed the synthetic magnitudes in our filter set, and found the
best-fitting template with a standard $\chi^2$ minimization, leaving
the normalization of the model magnitudes as a free parameter.

Within this general framework,
we introduce two improvements in the SED-fitting procedure, compared to
our previous papers:

1) We adopt three different parametrizations for the star-formation
history (SFH):

\begin{itemize}
\item
Exponentially declining laws (SFH$\propto exp(-t/\tau)$) with
timescale $\tau=0.1,0.3,0.6,1.0,2.0,3.0,5.0,9.0,15.0$\,Gyr
(``$\tau$-models'');
note that, at the redshifts of interest here, the models with large
$\tau$ (9,15 Gyr) are in practice equivalent to models with constant
star-formation rate, since the age of the Universe is much smaller
than the $\tau$-folding timescale.
\item
``Inverted-$\tau$'' law (SFH$\propto exp(+t/\tau)$) with the same
range of timescales as above.
\item
``Delayed'' star-formation history (SFH$\propto t^2/\tau \times
exp(-t/\tau)$) with $\tau$ going from 0.1 to 2.0 Gyr with a step of 0.1 Gyr.
This SFH law rises up to $t=2\tau$ and declines thereafter.
\end{itemize}

2) We include the contribution from nebular emission computed
following \cite{schaerer2009}. Briefly, in this model nebular emission is
linked to the amount of hydrogen-ionizing photons in the
stellar SED (\cite{schaerer1998}) assuming an escape fraction
$f_{esc}=0.0$ (Case B recombination).
The ionizing radiation is converted into nebular
continuum emission considering free-free, free-bound, and hydrogen two-photon
continuum emission, assuming an electron temperature $T_e = 10,000$\,K, an
electron density $N_e=100\,{\rm cm^{-3}}$, and a 10\% helium numerical
abundance relative to hydrogen. Hydrogen lines from the Lyman to the
Brackett series are included considering Case-B recombination, while
the relative line intensities of He and metals are taken, as a
function of metallicity, from \cite{anders2003}.

The SED-fitting has been performed separately for each of the
analytical SFHs listed above, both including and excluding nebular
emission: in the following we will refer to the masses obtained with
exponentially-declining models and no emission lines as the
``reference'' masses, for comparisons with previous studies.

Different spectral libraries could be adopted to derive stellar masses
for high-redshift galaxies. For example the libraries provided by
\cite{maraston05} and \cite{cb07} both attempt to take into account,
in slightly different ways, the contributions of evolved stars (in
particular during the TP-AGB phases), to the near-infrared emission
from galaxies.  The main contribution of TP-AGB stars occurs $\sim
0.5-2$\,Gyrs after an episode of star formation
(\cite{maraston05}). Since the age of the Universe at $z=3.5$ is 1.77
Gyr, we cannot neglect a priori the contribution of these peculiar
stars. However, recent estimates (\cite{santini12}) of the
differences between the masses derived through the BC03 and the M05 or
BC07 libraries indicate that there are only small dissimilarities between
the results from these stellar evolution codes at $z<4$, with the
largest differences found at $z\sim 2$. Considering also that these
models are undergoing revisions from their authors, we decided to use
only the BC03 library in this paper.

The derived galaxy masses for the whole GOODS-South+HUDF field are
shown as a function of redshift in Fig.\ref{mzcomp}. We also
indicate (purple big squares) galaxies with robust
spectroscopic redshifts in the range $3.5\le z\le 7.5$. Their positions in
the mass vs redshift diagram show that they are representative of the
overall distribution of galaxies in this redshift range.
In the interval $6.5\le z\le 7.5$ very few spectroscopic redshifts are
available, and our results rely mainly on the photometric redshifts.

\subsection{The Galaxy Stellar Mass Function estimate}

To compute the GSMF, we adopt two standard methods described in
\cite{fontana04,fontana06}, and \cite{santini12}. The
first is based on the non-parametric $1/V_{max}$ method by
\cite{schmidt68} and \cite{avnibahcall80}. For each
redshift and stellar mass bin, the total volume $V_{max}$ is derived
taking into account the magnitude limits in the different areas of the
survey, adopting the same technique normally used to compute the
luminosity functions. 
 
The second method is the STY (\cite{sandage79}) maximum likelihood
analysis assuming a \cite{Schechter1976} parametric
form. For each magnitude range considered, we compute the likelihood
to find the observed galaxies given the survey characteristics, the
various sources of incompleteness and the three parameters describing
the Schechter function. We then maximize the global likelihood for
the whole survey and find the best fit parameters for the mass
function. 

The major difference that needs to be introduced in computing a GSMF,
compared to a standard luminosity function, is an adequate handling of
the distribution of $M_*/L$ ratio of the galaxies in the sample. At a
given mass, some galaxies can be characterized, for example, by very
high mass-to-light ratios (due to large ages or high dust extinction -
e.g. \cite{dunlop07}) and thus can be much fainter than a more typical
blue star-forming and moderately obscured galaxy. To deal with this, we
adopt the technique, described in \cite{fontana04}, that allows us to
compute the fraction of objects lost because of their large
$M_{\ast}/L$, by measuring the actual distribution of $M_{\ast}/L$
immediately above the completeness limit in flux, and assuming that
this holds at slightly lower masses/fluxes.

\begin{figure}
\includegraphics[width=9cm,angle=0]{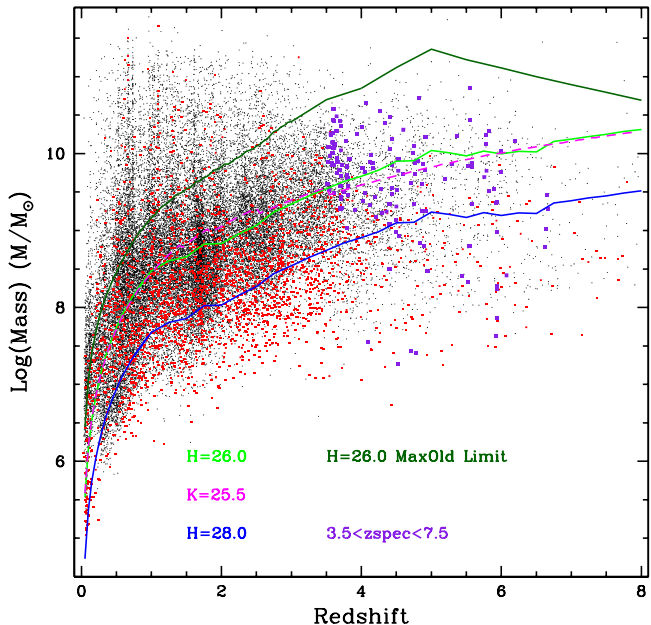}
\caption{
The black dots are
all the galaxies in the CANDELS GOODS-South field while the red dots
show those galaxies in the deepest region (HUDF). The purple big squares
indicate galaxies with robust spectroscopic redshift in the range
$3.5\le z\le 7.5$.
The strict completeness mass versus redshift for a magnitude limit of
$H_{160}=26.0$ (dark green). This curve has been derived from the
maximally old galaxies in our synthetic library, corresponding to a
formation redshift $z_{form}=20$, an $E(B-V)=0.1$, metallicity of
0.2${\rm Z_{\odot}}$ and a declining SFH with timescale
$\tau=0.1$\,Gyr. The light-green curve shows the
completeness-corrected limit corresponding to the shallower pointings
in the GOODS-South and UDS fields ($H_{160}=26.0$), while the blue
curve represents the corresponding limit for our deepest area, the
HUDF field ($H_{160}=28.0$), once the mass limit has been extended
taking into account the suitable correction for the $M^*/L$
distribution. For comparison, the magenta curve (dashed) is the
completeness limit in mass in the ERS field derived by
\cite{santini12} which corresponds to $K=25.5$.
}
\label{mzcomp}
\end{figure}

The derived galaxy masses, and the relevant mass completeness limits as 
a function of redshift are plotted in Fig.\ref{mzcomp}. This plot shows 
the strict completeness limit for $H=26$ (dark green line), but also shows that 
the minimum mass, at which the GSMF is computed, is in practice lower than 
this, because it takes into account the
appropriate correction factor for incompleteness. This limit is shown as a
light green curve in Fig.\ref{mzcomp} for the shallowest areas 
in GOODS-South and UDS, and by a blue line for the deepest HUDF pointing.
The black dots in Fig.\ref{mzcomp} show all the galaxies in the CANDELS
GOODS-South field while the red dots show those galaxies in the deepest
region (HUDF). The minimum mass above which the GSMF is computed is
well above the lowest mass galaxies detected in our survey, indicating that
this approach of extending the completeness mass is nonetheless robust.
The reader interested
in the technical issues on the calculation of completeness of the GSMF
is referred to \cite{fontana04} for all the details.

Based on an extrapolation of the mass-to-light distributions at
slightly brighter luminosities,
our survey, in the small but very deep HUDF field, can detect a galaxy
with mass $M=10^{9}\,{\rm M_{\odot}}$ at $z=4$, $M=2\times
10^{9}\,{\rm M_{\odot}}$ at $z=5$, $M=3\times 10^{9}\,{\rm M_{\odot}}$
at $z=6$ and $M=6\times 10^{9}\,{\rm M_{\odot}}$ at $z=7$. Thus, with
the CANDELS+HUDF survey, we can probe the GSMF at masses well below
the knee of the GSMF at $z=5-6$, with an acceptable precision, even at
relatively low stellar masses. We cannot exclude however the presence of
a rare population of very red dusty galaxies with
large masses at high-z. They would be characterized by extreme M/L
ratios, and consequently be too faint to be detected by the present
CANDELS survey.

\section{The uncertainties on the derivation of the Galaxy Stellar Mass
Function}

The uncertainties involved in the computation of the GSMF are numerous
and, unfortunately, difficult to estimate. Many of them, of course,
stem from the uncertainties involved in the evaluation of the stellar
masses of individual galaxies, and are hence larger than those
involved in the estimate of the luminosity function. To some extent,
they depend on conceptual aspects that have not been fully quantified
yet, such as the uncertainties on the actual star-formation histories of
galaxies, or the metallicity evolution. The impacts of other
effects, instead, like the intrinsic degeneracies of input models,
depends on the characteristics of the observations adopted and need to
be estimated carefully for any data set. We explore in this section
such uncertainties, focusing directly on the impact that they have on
the estimate of the GSMF. An analysis of the uncertainties of the
stellar masses estimated for individual galaxies in CANDELS is
presented with more details in two related papers (Mobasher et al., subm.,
Santini et al. subm.). These papers explore the
systematic and random uncertainties on a galaxy-by-galaxy basis. We
anticipate and use here some of their results, deferring to such papers
for a more detailed discussion, to show the effect that such
uncertainties have on the global GSMF.

We divide such uncertainties into two main categories: {\it random}
errors, i.e. those arising from photometric uncertainties, from the
errors on the redshift determination (quantified via the probability
distribution functions of photometric redshifts), and {\it systematic}
effects, such as the photometric redshift recipes (e.g. libraries,
obscuration law, IMF), the adoption of various SFHs, the inclusion of
nebular contributions in the SED fitting and from the field-to-field
variation, also known as Cosmic Variance. Last, we consider also the
effect of the AGN population on the high-redshift GSMF.

We anticipate here that the differences in the stellar mass estimates
using the different systematic variants mentioned above are small
($\Delta M/M<0.1$ dex), comparable to or lower than the statistical
uncertainties. We have also demonstrated that there are no strong 
trend of these systematic effects with either redshift or stellar mass $M$,
the only notable exception being the contribution of 
nebular lines and continuum at $z>4$. Masses
computed assuming strong emission lines are smaller than the fiducial
masses by 0.05-0.20 dex and depend on the redshift intervals where
strong lines enter the near-infared or IRAC filters (see Fig. 8 of
\cite{salmon13}). All these results will be shown in detail in
\cite{santini14} both for the UDS and GOODS-South fields.

\subsection{Sources of random errors}

We consider here the impact of different random effects (photometric
uncertainties and model degeneracies) on the GSMF estimation.

\subsubsection{The impact of photometric uncertainties and model degeneracies
on the stellar mass computation}

The derivation of the stellar mass for a galaxy is based on 
knowledge of its physical parameters (age, dust extinction,
metallicity). Since a given SED (even with the small photometric 
errors that we have in CANDELS) can be fitted with some combination of these
parameters, the stellar mass is inevitably uncertain due to unavoidable
degeneracies between them. This effect still exists even when the
spectroscopic redshift is known with high accuracy, and the uncertainties 
are obviously larger when a galaxy has only a photometric redshift.

To estimate the impact of these errors on the GSMF, we have carried
out a Monte Carlo simulation, specific to our data set. For each galaxy 
with a secure spectroscopic redshift we simply adopt it, while for any 
galaxy without a secure spectroscopic redshift, we extracted a random
redshift following the Bayesian probability distribution function
$PDF(z)$ computed as described in \cite{dahlen13}.
These probability distribution functions have been derived by combining
the $PDF(z)$ computed by six different groups within the CANDELS
collaboration. Before the bayesian combination, all the individual PDFs have
been slightly modified in order to recover the correct number of spectroscopic
redshifts within the errors, as discussed in detail by \cite{dahlen13}.
We then scan all
the models in the BC03 synthetic library at that redshift $z$
and compute the probability
distribution function of the stellar mass $PDF(M|z)$. Following this
distribution, we eventually extract a mass $M(z)$ which is compliant
with the observed SED of the specific object both in terms of
allowable input parameters and its CANDELS Bayesian photometric
redshift probability $PDF(z)$. For each object, the same procedure has
been repeated 1000
times, and used to estimate the 1~$\sigma$ uncertainty on its stellar
mass by computing $\Delta M=(M_{84}-M_{16})/2$, where $M_{16}$ and
$M_{84}$ are the 16th and 84th percentiles of the mass distribution,
respectively.

\begin{figure}
\includegraphics[width=9cm,angle=0]{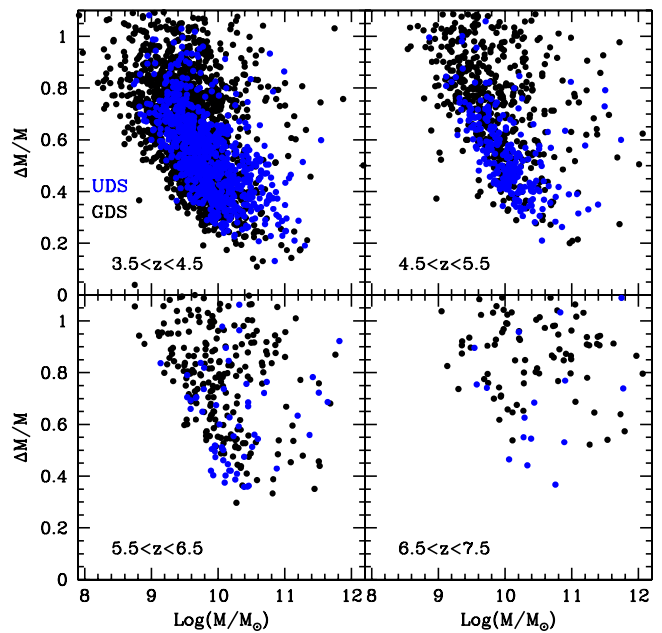}
\caption{The relative mass uncertainty, $\Delta M/M$, as a function of stellar
mass for different redshift bins from $z=4$ to $z=7$. Black points are for
galaxies in the GOODS-South field, while blue ones are for objects in the
CANDELS UDS region. The mass uncertainty is a strong function of the mass of the
galaxies, and as expected it is significantly poorer at higher redshifts.
}
\label{dmvsm}
\end{figure}

These values provide us with a clear indication of the level of
accuracy on stellar masses of {\it individual} galaxies, and are plotted in
Fig.\ref{dmvsm} as a function of the stellar mass in different
redshift bins from $z=4$ to $z=7$. At $z=4$ and for masses of the order of
$10^{10}\,{\rm M_{\odot}}$ the typical errors are of the order of 0.4\,dex,
and increase, as expected, towards higher redshifts and/or smaller masses
due to the increased photometric uncertainties and model
degeneracies. We do not find significant differences between the
$\Delta M/M$ computed for the GOODS-South or the UDS field.
As we will show in Appendix B, the uncertainties in mass we have derived
are not Gaussian
and, especially at high redshifts and low masses, they are asymmetric, with
a trend towards lower masses in general.

It is important to note that this level of accuracy represents a distinct
improvement over previous surveys. We can make a direct comparison
with our previous data, where we used a comparable approach to estimate the
uncertainties. For comparison, a similar accuracy (0.4\,dex)
was reached at lower redshift ($z<3$) and brighter magnitudes ($K\le
23.5$) in the original GOODS-MUSIC data set (\cite{fontana06}), that
used shallower ground based NIR imaging in the $J,H,K$ bands and the first
maps by Spitzer on the GOODS-South field (\cite{grazian06a}).
In \cite{santini12}, using data from the ERS
survey by {\it HST}, we reached a $\Delta M/M\sim 0.2-0.3$ but only at $z\le 4$.

This Monte Carlo simulation has been used to estimate the resulting
uncertainty in the GSMF. Using the simulations described above we
have obtained 1000 different realizations of the GSMF, and used them
to evaluate the scatter in the number density of each
bin in mass. This source of error will be labeled `MCsim' and it will be
compared with other systematic errors in below.

\subsection{Systematic errors}

We consider here the impact of different systematic effects
(photometric redshifts, star-formation history, nebular contribution,
cosmic variance,
AGN contamination) on the GSMF estimation. For simplicity, we do not
mention here the results for individual galaxies (we refer the reader
to Mobasher et al., subm., Santini et al. subm. for
full details) but only the effect on the GSMF. We therefore compute
different GSMFs with the various assumptions, and present the scatter
measured directly on their values.

\subsubsection{Photometric redshifts}

We re-emphasize that the photometric redshifts available for our
catalogues have been produced with a Bayesian average of six different
photometric-redshift solutions obtained with different codes and
techniques. The availability of completely independent photometric
redshift solutions gives us a unique opportunity to verify how
different recipes for photo-$z$ yield {\it systematic} differences in
the final estimate of the GSMF. We note that this effect is different
from what we have presented in the previous section, where we have
included the effect of the uncertainty in the redshift as estimated
{\it internally} to a given technique (in this case our Bayesian
average). The analysis that we describe here is useful for estimating
the extent to which the existing differences between published
determinations of the GSMF can be ascribed to a scatter induced by the
various photo-$z$ techniques.

\begin{figure*}
\centering
\includegraphics[width=14cm,angle=-90]{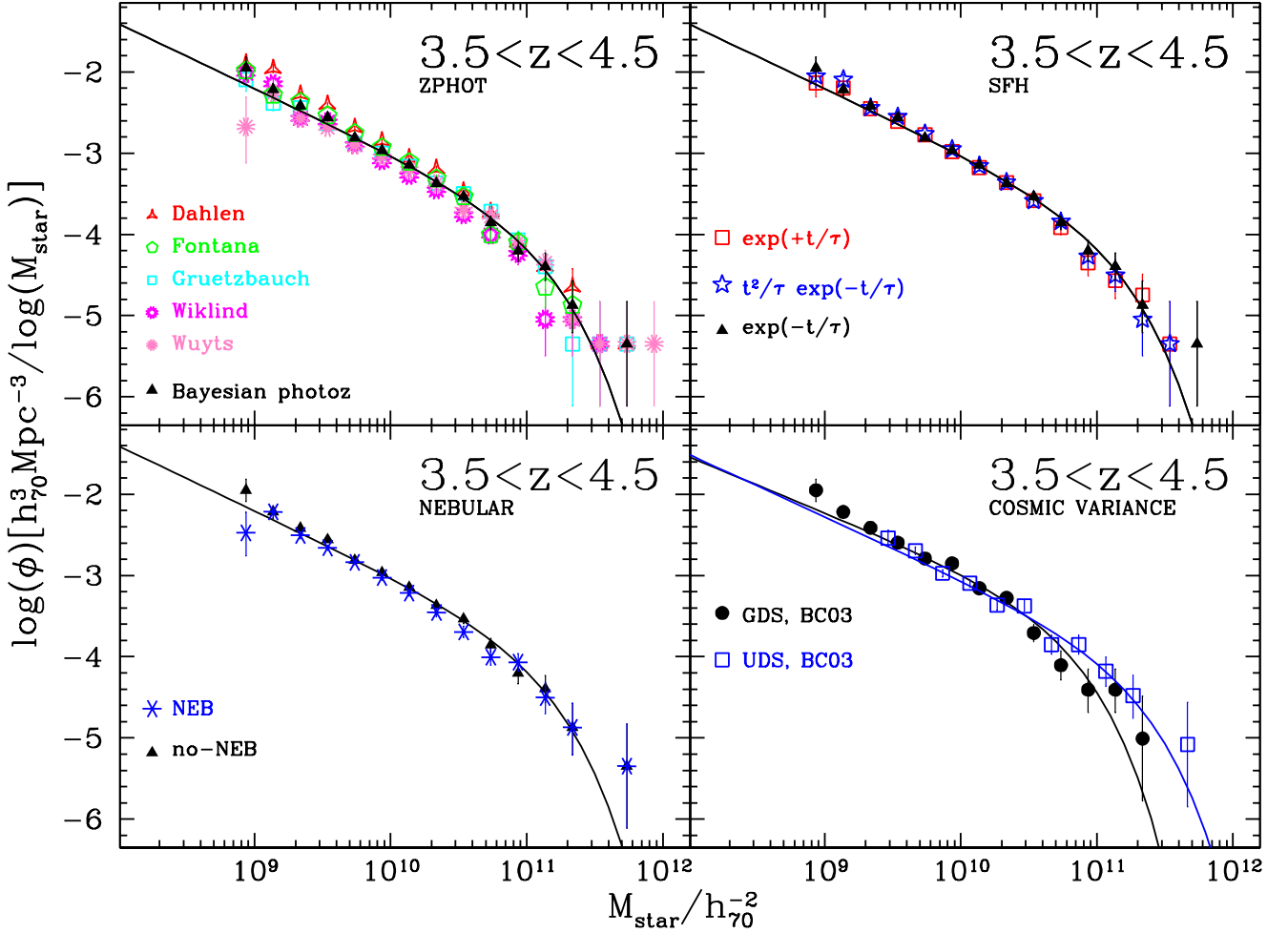}
\caption{{\em Top left: Comparison of the GSMF obtained with different
photometric redshift recipes.} The stellar mass function of
galaxies at $3.5\le z\le 4.5$ in the CANDELS GOODS-South and UDS
fields derived using the Bayesian photometric redshifts is shown by
the black triangles (non parametric $1/V_{max}$) and the solid
continuous curve (parametric STY Maximum Likelihood). The error bars
show the Poissonian uncertainties of each point. The red
triangles, green stars, cyan triangles, magenta and pink asterisks
show the GSMFs obtained using the individual photometric redshifts of
different realizations that have been used to derive the Bayesian
photo-$z$ described in \cite{dahlen13}.
{\em Top right: Comparison of the GSMF obtained with different
star-formation histories.} The stellar mass function of galaxies at
$3.5\le z\le 4.5$ in the CANDELS GOODS-South and UDS fields assuming
BC03 exponentially-declining SFHs is shown by the black triangles and
the solid continuous curve. The red squares and the blue stars show
the GSMFs derived using exponentially increasing and a delayed
SFH, respectively. All these star formation
histories have been tested without the nebular contribution.
{\em Bottom left: Comparison of the GSMF with and without a nebular
contribution.} The stellar mass function of galaxies at $3.5\le
z\le 4.5$ in the CANDELS GOODS-South and UDS fields with BC03
exponentially-declining SFHs and no nebular contribution is shown by
the black triangles and the solid continuous curve. The blue asterisks
show the GSMF derived using BC03 and exponentially-declining SFHs but
this time including allowance for the contribution of nebular lines
and nebular continuum (NEB label).
{\em Bottom right: Comparison of the GSMF in individual fields.} The
stellar mass function derived for galaxies at $3.5\le z\le 4.5$ in the
CANDELS UDS field (empty blue squares) as compared with that derived
from GOODS-South (filled black circles).
}
\label{sumcmp}
\end{figure*}

To achieve this, we have computed the GSMF with the individual recipes for
photometric redshift used to assemble the average photo-$z$ used here
(\cite{dahlen13}). To simplify the comparison, we have estimated stellar masses
simply using BC03 models with no emission lines and standard
exponentially declining histories as a baseline. The results are shown in
Fig.\ref{sumcmp} (top-left) for the redshift range $3.5<z<4.5$; the
GSMFs for the other redshift intervals are shown in the appendix A.

Comparing the individual mass functions, it is possible to notice
systematic differences, which at low redshift ($3.5<z<4.5$) and at
intermediate stellar masses ($M\sim 10^{10}M_{\odot}$) are comparable
to or slightly larger than the Poissonian errors (represented by the error
bars in the plot). The scatter between individual GSMFs is enhanced
at the high-mass end, where the photo-$z$ leakage of even a few
galaxies into the highest redshift bins may cause a significant
increase in the estimated number density. The Bayesian photo-$z$ are
less subject to this redshift leakage by construction. In addition,
the large scatter between different photometric redshift realizations
in the exponential tail can be due also to the low number statistics,
since at the massive end only a few galaxies contribute to the
GSMF. Galaxies with high stellar masses, indeed, could be highly
obscured by dust or characterized by an old stellar population, and in
this case the UV optical magnitudes are expected to be faint, with a
consequent relatively low accuracy of the photometric redshift
solutions.

The low-mass-end slope of the GSMF, instead, is less sensitive to the
photo-$z$ recipes adopted, and all the different photometric redshift
methods confirm that the GSMF is apparently steepening from $z=4$ to
$z=7$, as shown in Fig.\ref{rec}.

\subsubsection{Star formation histories}

We have investigated also the effect of different parametrizations of
the Star Formation History (SFH) on the mass function analysis. As
described in Sect. \ref{sec:stellarmasses}, we have considered separately
the classical exponential declining model, the exponentially
increasing SFH (\cite{maraston10,pforr12}) and the ``delayed'' SFH
that increases at early ages and then shows a declining phase at later
epochs. The effect of all these star formation histories on the GSMF
has been tested without the nebular contribution at this stage.

Fig.\ref{sumcmp} (top-right) shows the different mass function estimates
assuming different star-formation histories at $3.5<z<4.5$.
A similar plot for the other redshift range is shown in the appendix A.
Small differences can be noticed
at the massive tail of the distribution ($M\ge M^*$), while the slope of the
low-mass end of the GSMF is stable against the adoption of different SFHs.
This is consistent with the results of \cite{mobasher14} and
\cite{santini14}, where it is shown that the stellar mass parameter is
practically insensitive to the choice of the adopted SFHs, due to degeneracies
with both age and dust extinction.
We thus confirm that the choice of the SFH does not strongly
influence the form of the inferred GSMF at high redshift.

\subsubsection{The impact of nebular lines and nebular continuum}

We have considered also the impact of nebular emission (both lines and
continuum) on the stellar mass and GSMF derivation. To explore this
we have adopted here the approach taken by \cite{schaerer2009}, followed also by
\cite{duncan14}. This involves deriving the production rate of
ionizing photons by integrating the intrinsic BC03 template at
$\lambda<912\AA$ rest frame, and then converting ionizing flux into
both nebular lines and continuum. These models assume a constant
temperature and electron density, which is probably a somewhat rough
approximation, but nonetheless gives realistic results,
as shown in \cite{castellano14} for a sample of galaxies at $z=3-4$
(taken mainly from the AMAZE sample of \cite{amaze} and \cite{troncoso14}).

Fig.\ref{sumcmp} (bottom-left) shows the impact of the nebular
contribution to the GSMF estimate at $3.5<z<4.5$. A similar plot for
the other redshift ranges is shown in the appendix A. At $z\simeq 4$
we find agreement with the results of \cite{stark13} and
\cite{salmon13}, indicating that the differences in the stellar masses
computed with and without the nebular contribution are less than
0.1-0.2 dex. The relatively low contribution of nebular emission on the mass
determination at high redshift can be due to the large FWHM in
wavelength of the IRAC filters ($\sim 0.8-1.0\mu$m). At higher
redshifts ($z>4.5$) the differences increases slightly, since stellar
masses are systematically shifted lower, due to the larger relative
contribution of emission lines in the {\it Spitzer} bands, but is in
any case within 0.2 dex. This result is consistent with similar
results obtained by \cite{duncan14} and \cite{salmon13}, who have also
attempted to take into account the potential nebular contribution.

Our results are also consistent with those obtained by \cite{stark13} 
at $3.5<z<6.5$. At $6.5<z<7.5$ they find a larger offset in stellar masses of
0.3 dex; this difference could be due to the different method adopted,
since their estimate at these large redshifts has been derived
{\it assuming} the same EW distribution of the $H\alpha$ line emission
inferred at $3.8<z<5.0$.

In summary, we find that the nebular contribution
does not alter dramatically the shape of the GSMF, even at very high redshift 
($z=6-7$). There is, as expected, possibly a slight systematic effect towards 
lower number densities at a given mass when the nebular contribution is 
allowed in the fitting, but this trend is within the uncertainties of the
GSMF and always less than $0.1-0.2$\,dex in $\log(\Phi)$.
The difference in the faint end slope $\Delta\alpha$ computed on the
Mass Functions with and without nebular contribution at
$z\sim 7$ is $\sim 0.04$, confirming the robustness of the GSMF against
this systematic effect.

\subsubsection{The Cosmic Variance}

Another source of uncertainty in the GSMF estimation is the field-to-field
variation, also known as cosmic variance. Despite the relatively large
area covered by the CANDELS survey at an unprecedented depth, the volume
sampled by deep HST observations is not larger than the possible scales of
over-densities and under-densities at those redshifts 
(\cite{ouchi09}).
For example, \cite{lee2012} found strong cosmic variance between the
number counts of LBGs at $z=4$ and $z=5$ between the GOODS-South and
GOODS-North fields. Cosmic variance can also be an important source of
the scatter observed so far in the various estimates of the GSMF at
high redshift, especially at the high-mass end.

To make a first-order estimate of the amplitude of the cosmic
variance, we compare the GSMF for the two fields (GOODS-South and UDS)
separately in Fig.\ref{sumcmp} (bottom-right) for the redshift
interval $3.5<z<4.5$. The same plot for the higher redshift
bins is shown in the appendix A. As expected, the biggest
uncertainties/differences are at large masses ($M\ge
10^{10.6}M_{\odot}$ at $z=4$). We find a similar trend at higher
redshifts, as shown also in the appendix A. At z=7 the difference
between the best fit of the GSMF in the UDS and GOODS-South field is
$\Delta\alpha\sim 0.037$, but the value of $M^\ast$ is significantly different
for the two fields, $logM^\ast_{GDS}=9.64$ against $logM^\ast_{UDS}=11.95$.
This indicates that we need
larger areas to beat down the cosmic variance: the completion of the
CANDELS survey and the availability of other deep fields being
observed with {\it HST} will probably provide the necessary
combination of depth and area to overcome this limitation.

\subsubsection{The presence of AGN in the input CANDELS sample}

The presence of AGN in principle can alter the stellar mass derivation
since the radiation from the active super-massive black holes, if
ignored, is usually converted into stellar masses adopting pure
stellar libraries.

Adopting the same technique of \cite{fontana06}, we exclude here all
the spectroscopically-confirmed AGN (both type 1 and 2) and the
luminous hard X-ray detected objects from the parent CANDELS sample
(\cite{xue11}). It is worth noting that in the GOODS-South field the
identification of additional high-redshift AGN is also possible thanks
to the variability studies and the wealth of multi-wavelength data
available (\cite{trevese94,villforth}), from ultra-deep X-ray imaging
by Chandra to the mid- and far-IR ({\it Spitzer, Herschel}).  In the
GOODS-South field we thus exclude 22 AGN at $3.5\le z\le 7.5$ from the
parent sample ($\sim 1\%$). The removal of AGN in the UDS field is
not as trivial as in the GOODS-South case, since such a photometric
database is not available or is shallower, at the moment. Moreover, at
$z>6$ the identification of AGN is made more difficult due to flux
limits on the X-ray and on optical spectroscopic identification. For
this reason, investigating the contribution of AGNs to the GSMF
estimation is important.

\begin{figure}
\includegraphics[width=7cm,angle=-90]{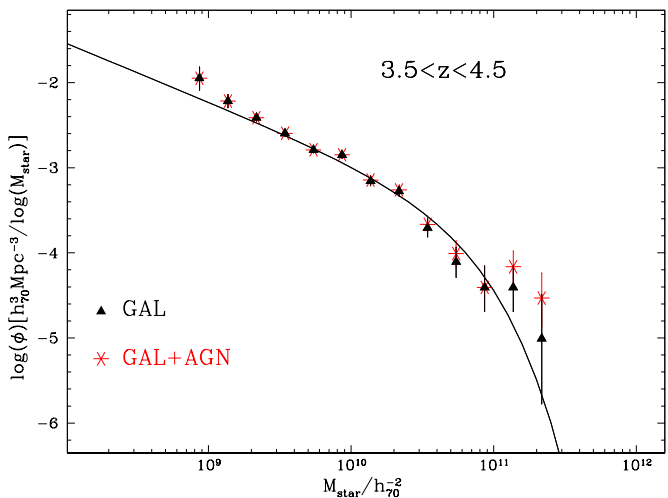}
\caption{{\em Comparison of the GSMF with and without AGN.}
The stellar mass function of galaxies at $3.5\le z\le 4.5$ in the CANDELS
GOODS-South field without AGN
(filled black triangles) and with the AGN included (red asterisks).
The error bars show the Poissonian uncertainties of each point.
The stellar masses for the AGN have been derived using the same technique
adopted for normal galaxies. The solid continuous curve shows the
Schechter function derived through a parametric STY Maximum Likelihood fit
of the GSMF without AGN.
}
\label{agnz4}
\end{figure}

In Fig.\ref{agnz4} we compare the galaxy-only GSMF with that which is
derived deliberately retaining all known AGN in the sample (but
estimating their stellar masses using pure stellar libraries).  We
find that at $3.5<z<4.5$ the two are almost identical at $M\le
10^{11}\,{\rm M_{\odot}}$, and within the uncertainties for higher
masses. The plot which summarizes the comparison at all redshifts can
be found in the appendix A. From these checks, we can conclude that at
masses lower than $10^{11}\,{\rm M_{\odot}}$ the role of AGN is
negligible, and at the massive tail they are introducing changes that
are within the uncertainties. This is as expected, since bright AGN
tend to populate the center of massive galaxies.

\subsection{Comparison of different sources of uncertainties}

Armed with a full characterization of the random and systematic
effects, we are now in a position to compare them and assess the
overall reliability of the GSMF. For all the random and systematic
effects described above, we have estimated the r.m.s. of the different
measured densities $\Phi(M)$ of the GSMF in each mass bin.

\begin{figure}
\includegraphics[width=9cm,angle=0]{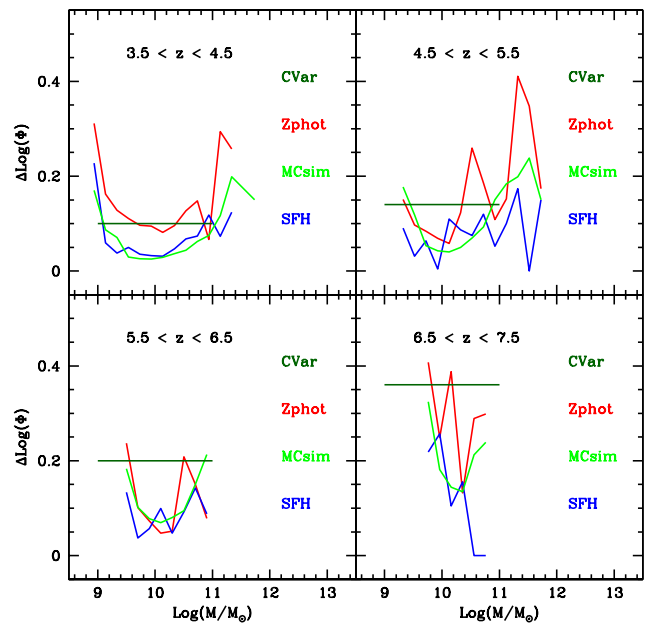}
\caption{The various uncertainties in $Log(\Phi)$ as a function of stellar
mass for different redshift bins from $z=4$ to $z=7$. Blue curves (SFH) indicate
the uncertainties due to the different SFHs adopted in this work. Green
curves (MCsim) are related to the Monte Carlo simulations described in 
paragraph 4.1. The red lines (Zphot) indicate the error introduced by
different photometric-redshift solutions. Finally, the dark-green lines (CVar)
indicate the error due to Cosmic Variance.
}
\label{errmf}
\end{figure}

As far as the random errors are concerned, we compute the
uncertainties on the GSMF due to the combined probability distribution
functions in photometric redshift and stellar mass $PDF(z,M)$ using
the 1000 Monte Carlo simulations described in the previous paragraph.
The uncertainties are then derived measuring the r.m.s. of these GSMFs
and are indicated in Fig.\ref{errmf} with the label ``MCsim'' (Monte
Carlo simulations).  For the cosmic variance effect we do not rely on
the r.m.s. computed only on two fields, but we used the ``Cosmic
Variance'' tool\footnote{\em
http://casa.colorado.edu/$\sim$trenti/CosmicVariance.html} provided
by \cite{trenti} using as input parameters the number of galaxies
observed in our two fields.

Similarly, we have computed the GSMF that results from the adoption of
different methodologies, such as photometric-redshift recipes,
star-formation histories (exponentially declining, exponentially
rising, delayed) and the nebular model.

These errors are compared in Fig.\ref{errmf}, that shows the different
uncertainties in $Log(\Phi)$ as a function of stellar mass for
different redshift bins from $z=4$ to $z=7$. Table \ref{tabuncmf}
summarizes the median uncertainties in all the mass bins for
$3.5<z<7.5$, due to the photometric redshifts, the star-formation
histories and the nebular contribution (SFH+NEB), the Monte Carlo
simulations and the field-to-field uncertainties (CVar).

\begin{table}
\caption{Median uncertainties on the GSMF due to photometric redshifts,
Monte Carlo simulations or SFH+NEB}
\label{tabuncmf}
\centering
\begin{tabular}{c | c c c r}
\hline\hline
Redshift & $\sigma_{log(\Phi)}$ & $\sigma_{log(\Phi)}$ & $\sigma_{log(\Phi)}$ \\
 & photo-$z$ & SFH/NEB & ``MCsim'' & $CVar$ \\
\hline
$3.5<z<4.5$ &  0.13 & 0.06 & 0.07 & 0.10 \\
$4.5<z<5.5$ &  0.15 & 0.09 & 0.12 & 0.14 \\
$5.5<z<6.5$ &  0.12 & 0.09 & 0.11 & 0.20 \\
$6.5<z<7.5$ &  0.29 & 0.13 & 0.20 & 0.36 \\
\hline
\end{tabular}
\\
$CVar$ is the cosmic variance error of the
GOODS-South+UDS fields computed with the recipes of \cite{trenti}.
\end{table}

This comparison shows that the errors in the GSMF are relatively
small, of the order of 10-20\% in most of the bins, when the
different sources of uncertainties are individually taken into
account (photometric redshift recipes, star
formation histories, nebular contribution). In general, we note
that errors tend to be larger at the high-mass end (where the
number of objects is small) and at small masses (where objects are
faint and hence more susceptible to errors), and smaller at intermediate
masses, where the statistics are better and photometry is still highly
reliable.

It is also clear that the leading source of errors is the adoption of a
specific recipe for the computation of the photometric redshifts. This
effect dominates over those due to the different parametrizations of
the SFH or the adoption of the nebular contribution. We remark that
this test has been performed here for the first time, to our
knowledge, thanks to the various recipes developed and compared within
the CANDELS collaboration. This test is different from the evaluation
of the effects of noise in the redshift estimate internal to a given
technique, that we have also performed and that is labeled `MCsim' 
(and which was already included in previous analyses;
e.g. \cite{fontana06,marchesini09}), but it is instead related to the
systematic differences that may arise when different techniques are
adopted.

Our check suggests that a significant contribution to the observed
scatter among the various GSMF presented in the literature, at least
at high redshift, can be ascribed to the different photometric redshift
techniques adopted.

We also note that the typical error due to the uncertainties in
photo-$z$ (`MCsim') in our survey is somewhat lower than in previous
surveys thanks to the improved quality of the photometry and to the
adoption of a Bayesian approach that improves the accuracy and leads
to narrower $PDF(M|z)$ distributions. Our decision to use the Bayesian
photo-$z$ by \cite{dahlen13} allows us to reduce the uncertainties in
the GSMF at high redshift by $\simeq 0.1-0.3$\,dex.

The other fundamental ingredient to constrain the uncertainties of the
GSMF is the limitation of the cosmic variance effect, and this is possible
adopting observational
strategies tailored to maximize the efficiency of the surveys.
From Table \ref{tabuncmf} it is clear that the uncertainty on the photometric
redshifts and the cosmic variance effect of the CANDELS GOODS-South+UDS
fields are comparable. Larger areas or multiple fields would be essential
in the future to further beat down the field-to-field variations.

The previous tests have allowed us to quantify the effects of different
``ingredients'' on the mass function, namely photometric redshifts, star
formation histories and nebular contribution. However, these have been
considered separately, i.e. varying only one parameter at time and
checking for its effects on the GSMF, as shown in Table \ref{tabuncmf} and
in Fig.\ref{errmf}. To quantify the co-variance between
these three main ingredients, we have considered all their possible
combinations,
varying simultaneously the five recipes for photo-z including
the bayesian solution, the three SFHs
and adding or neglecting the nebular contribution. We have thus produced
36 different GSMFs. As a consequence, the total variance of these
determinations gives a quantitative estimate for the co-variance between
these three ingredients. We find that the scatter of $Log(\Phi)$, obtained
from the 36 different mass function realizations,
is very similar to the square-root combination of the individual
variances due to photo-z, SFHs and nebular emissions.
We can thus conclude that the co-variance terms between
photo-z, SFHs and nebular contribution are negligible and 
that they can be effectively factorized in different independent variances.

Here we did not consider other sources of variance on the mass function
determinations, i.e. mass-variable IMF, different libraries for simple stellar
population synthesis, non-uniform dust screen
models, different dust laws, and other parameters influencing the stellar
mass derivation through SED fitting. Taking into account all these
parameters the variance on GSMF can be even larger than what we find here.
We thus advertise the reader that the uncertainties summarised in Table
\ref{tabuncmf} and in Fig.\ref{errmf} are only an underestimation of the
true total error budget on the GSMF at high-z.


\section{The constituents of the GSMF at high redshift}

\begin{figure*}
\centering
\includegraphics[width=14cm,angle=-90]{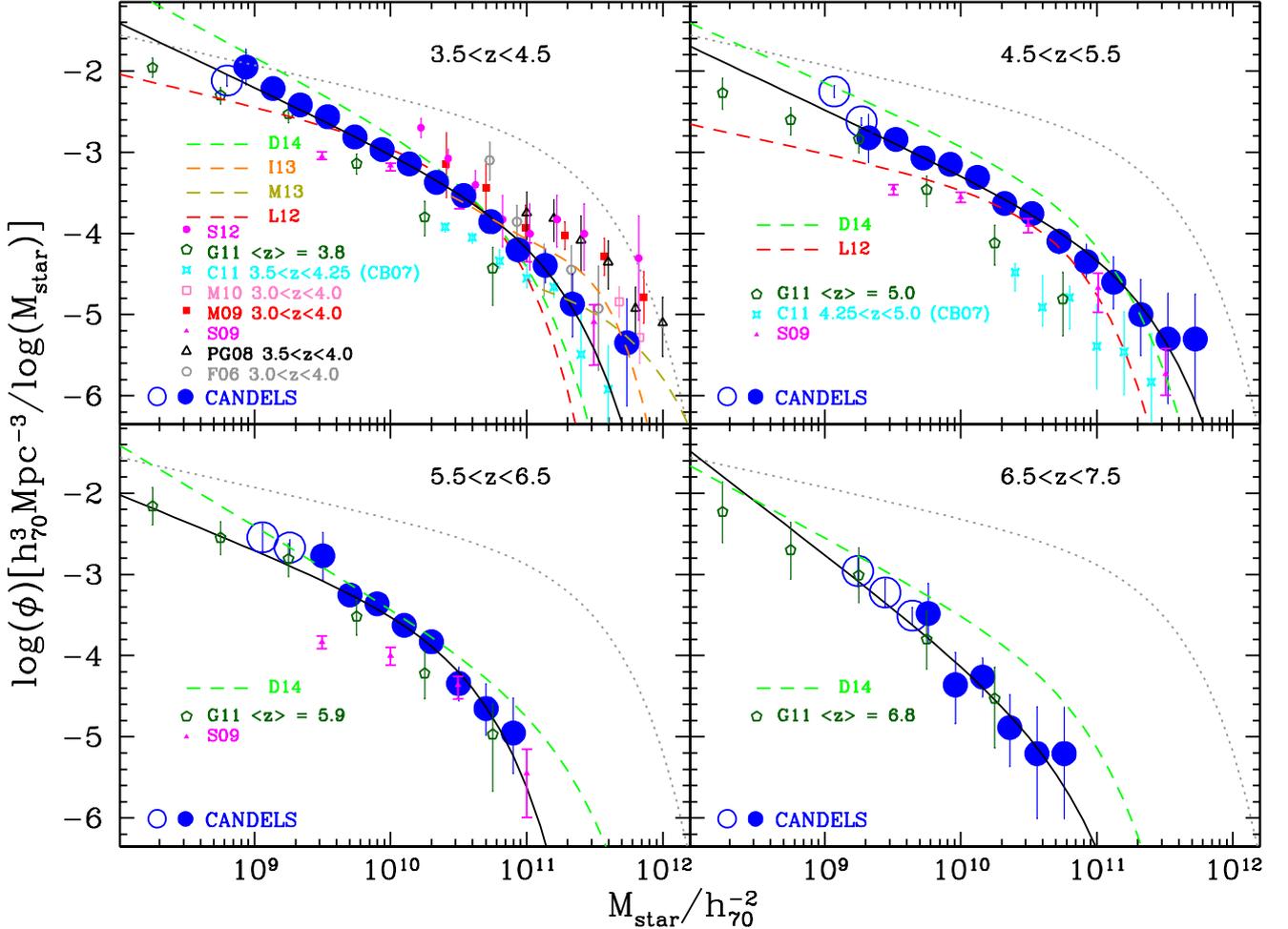}
\caption{
The stellar mass function of galaxies at $3.5\le z\le 7.5$ in the
CANDELS UDS and GOODS-South fields (blue filled and open circles).
The error bars
show the Poissonian uncertainties of each point with the errors derived
through the Monte Carlo simulations added in quadrature.
The masses
are derived using the BC03 libraries with exponentially-declining 
star-formation histories, and without any contribution from nebular lines
or continuum. AGN were not included in the present sample. The
dotted lines indicate the GSMF at $z=0.6$ in the UDS and GOODS-South
fields. The dark-green pentagons show the mass function derived by
\cite{gonzalez2011}~(G11), while the cyan stars indicate the result of
\cite{caputi11}~(C11), which was obtained with a different stellar library
(\cite{cb07}) that includes a stronger contribution from TP-AGB stars. 
The black triangles are from \cite{pg08}~(PG08), the red (empty and filled)
squares from \cite{marchesini09}~(M09) and
\cite{marchesini10}~(M10), respectively. The magenta points are the
GSMF of \cite{santini12}~(S12). The grey circles come from
\cite{fontana06}~(F06) while the magenta triangles are from
\cite{stark09}~(S09). The red, orange, dark-yellow and green dashed lines
show the best
fit GSMFs of \cite{lee2012}~(L12), \cite{ilbert13}~(I13),
\cite{muzzin13}~(M13) and \cite{duncan14}~(D14), respectively.
All the mass functions have been converted to a Salpeter IMF for
comparison.
The solid continuous curves show the Schechter function
derived through a parametric STY Maximum Likelihood fit.
}
\label{mfallhz}
\end{figure*}

\subsection{The CANDELS GSMF at z=4-7}

Fig.\ref{mfallhz} shows the GSMF at $3.5\le z\le 7.5$ derived by
combining the CANDELS GOODS-South and UDS fields (filled and open blue
circles). Most of these GSMFs (i.e. those represented by filled
circles) have been derived adopting our ``baseline'' mass estimation;
i.e. SED fitting to the whole photometry using BC03 models with
exponentially-declining SFHs, no nebular contribution,
excluding AGN from the parent sample, and adopting the Bayesian
photometric redshifts. We extend the GSMF at the very low mass end (as
shown by open circles) by converting UV luminosity into stellar mass adopting 
a constant $M/L$ ratio. The procedure will be described in
detail in the Sect. 5.2 and Sect.\ref{GSMFevol}, where we will
present the consequences and lessons learned from our new estimate of
the GSMF. We first compare our new determination of the GSMF with 
results from previous studies, and discuss
plausible origins for the discrepancies that we find.

As can be clearly seen from Fig.\ref{mfallhz}, the various estimates
of the GSMF at $z=4$, and in particular at $M\ge 10^{11}\,{\rm
M_{\odot}}$ differ quite dramatically. For example, if we compare
our results with the GSMF derived by \cite{stark09}, \cite{lee2012} or
\cite{gonzalez2011}, we find an excess of galaxies at the high-mass
end and a steeper slope at low masses. The high mass range is
particularly sensitive to the systematics that we described above,
such as cosmic variance and different recipes for photometric
redshifts. In addition, the photometric quality of the various
data-sets used in different studies varies significantly, and can
further contribute to the observed scatter. Indeed, the basic
selection wavelength used can be very different. For instance,
\cite{caputi11} adopted a catalogue selected directly from the {\it
Spitzer} images at 4.5$\mu$m, rather than utilising an $H$-band
selected catalogue as adopted here or in \cite{santini12}. Meanwhile,
bluer selection bands were used by \cite{stark09} and \cite{lee2012}
who undertook their primary galaxy selection in the $i_{775}$ and
$z_{850}$ ACS bands respectively (sampling the UV rest-frame
wavelengths at $z\ge4$). A more detailed discussion on these
differences, especially on the massive side of the GSMFs,
has been carried out in Section 5.3.

Although we are using deeper WFC3/IR data, 
the \cite{gonzalez2011} GSMFs extend to lower masses
than our mass function determinations. This is because 
the \cite{gonzalez2011} GSMF estimate is based on the UV luminosity 
function, rather than on a directly mass-selected sample.
In the next section we will discuss these differences in more detail, and 
will also investigate the nature of the galaxies at the high-mass end
and the relation between mass and UV light.

At $z\geq5$ the number of available GSMF is much smaller, and the
general agreement improves. We suspect that this is due
to the fact that, in general, the surveys adopted to estimate the GSMF
at extreme redshifts are of superior quality, and that the strong
signature provided by the IGM absorption makes the photo-$z$ more
robust in this redshift range. The main discrepancy is found again
with the \cite{caputi11} GSMF at $z\simeq 5$, and again we suspect
that the different selection criterion may have played a role.
At $z\simeq 7$ our GSMF slightly differs from the \cite{duncan14} one
at $M\sim 3\times 10^{10}\, {\rm M_{\odot}}$,
but this can be due to the low number statistics of the adopted samples.
Nonetheless, it is worth noting that the GSMFs at $z\geq5$ shown in
Fig.\ref{mfallhz} (by \cite{stark09,gonzalez2011,lee2012,duncan14})
have been derived from similar photometric databases
(including the GOODS-South field), so the cosmic variance scatter
may not be a dominant effect in this case.

\subsection{The Mass-to-light ratio of galaxies at $z\ge 3.5$}

As already mentioned, most previous attempts to derive the GSMF at 
very high redshift ($z>3$)
have been carried out through the conversion of rest-frame UV light into
masses (\cite{gonzalez2011}), assuming a tight correlation
between SFR and the stellar masses of galaxies observed at $z=4$.
Our SED fitting procedure is not only providing the stellar masses of the
analysed sample, but it also gives the absolute magnitudes at different
rest-frame wavelengths. We have therefore derived the mass-to-light ratio
at 1400 \AA ~rest frame for galaxies at $3.5<z<4.5$ in order to 
compare it with previously-derived relations from other studies.

\begin{figure}[!h]
\centering
\includegraphics[width=9cm,angle=0]{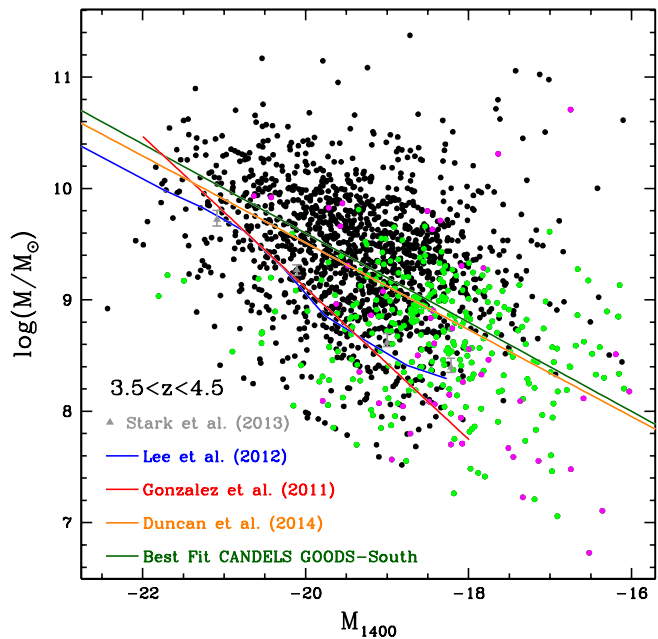}
\caption{This plot shows the stellar mass versus UV absolute magnitude
$M_{1400}$ for galaxies at $3.5<z<4.5$ in the GOODS-South field.
The blue solid line is the
relation (at 1700 \AA ~rest frame) found by \cite{lee2012} for LBGs
selected with the $B-V$ vs $V-z$ colour criterion and a S/N ratio in
the $z_{850}$ band greater than 6. The red solid line represents the
relation as derived by \cite{gonzalez2011} at $z=4$, for a similar
rest-frame wavelength of 1500\,\AA. The grey triangles show the $M/L$
relation derived by \cite{stark13}. The dark-green line is the
best fit to our own results assuming a constant mass-to-light ratio,
or equivalently a slope $-0.4$ between stellar mass and absolute
magnitude. The orange line is the best fit of \cite{duncan14}.
All the relations have been converted to a Salpeter IMF for comparison.
Magenta points show galaxies undetected in the deep
$K$-band Hawk-I imaging, while green dots represent objects not
detected in 3.6 and 4.5\,$\mu$m in the {\it Spitzer} SEDS imaging.
}
\label{masstolight}
\end{figure}

Fig.\ref{masstolight} shows the resulting relation between the stellar
mass $M$ and the UV-rest frame luminosity $L_{1400}$ for galaxies at
$3.5<z<4.5$ in the CANDELS GOODS-South field. Green points represent
the objects that, although too faint to be detected in the 3.6$\mu$m
and 4.5$\mu$m {\it Spitzer} bands, are instead detected in the deep
$K$-band HUGS observations. At $z\sim 4$ the $K$-band enables a direct
measurement of the rest-frame luminosity beyond the Balmer break,
where the bulk of the light from ordinary stars is manifest. This is
crucial for a robust estimate of stellar mass, and so there is no
doubt that the availability of these deep $K$-band images represents a
major improvement of the data now available in the CANDELS fields for
high-redshift mass determinations.

At $3.5<z<4.5$ a linear regression to the observed data yields the
relation $\log M=-0.4 \times M_{UV} + 1.6$, shown by a thick green
line in Fig.\ref{masstolight}. This relation implies that, albeit with 
large scatter, our data are consistent with a constant $M/L$
ratio. This equation is similar to the one deduced by \cite{duncan14}, but
with a slightly higher normalization. The trend of constant
mass-to-light ratio is valid also at higher redshifts, corroborating
the results of similar works on the CANDELS survey
(\cite{salmon13,duncan14}). The large scatter that we observe is
partly due to noise (in the photometry and in the derived rest-frame
quantities) and partly to a genuine scatter of the $M/L$ ratio in high
redshift galaxies, but the relative weight of the two aspects is
however difficult to quantify. We will discuss further below how
this impacts the resulting GSMF.

When compared with previous surveys, our results are in agreement 
at bright UV magnitudes, but become progressively more different 
at fainter UV luminosities. As we show
in Fig.\ref{masstolight}, the relations found by \cite{lee2012} and
\cite{gonzalez2011} do not reproduce the slope of the $M/L$ 
relation derived here.

In particular, the blue solid line in Fig.\ref{masstolight} is the
relation at 1700 \AA ~rest-frame found by \cite{lee2012} for LBGs
selected with the $B-V$ vs $V-z$ colour criterion and S/N ratio in the
$z_{850}$ band greater than 6, and this can be seen to be more
consistent with the lower envelope of our data than with our own
average $M/L$ relationship. This also appears to be the case for the
relation derived by \cite{gonzalez2011} (red solid line) who also used
samples of LBGs selected at $z=4$ (i.e. at a rest-frame wavelength of
1500\,\AA).

We first tried to reproduce the trend observed by \cite{gonzalez2011}
using only LBGs selected via the $B-V$ vs $V-z$ colour-colour
criterion, or fitting the masses with models of constant star
formation histories and/or solar metallicity, but we find that our
data points are always best fitted by a constant M/L
relation. Following the example of \cite{mclure11}, we explored also
synthetic libraries with constant star-formation histories and no
extinction, but the results are similar to our baseline model,
indicating (on average) a constant mass-to-light ratio.

A possible explanation for the differences could be the fact that the
relation between mass and light deduced by \cite{gonzalez2011} (their
Fig.1) appears to be driven - and possibly tilted - by the points at
lower masses that are derived from galaxies that are essentially
undetected ($S/N\le 2$) in the IRAC 3.6$\mu$m band. In our case,
instead, the estimates at low luminosity benefit from the combination
of the new deep HUGS Hawk-I $K$-band photometry (\cite{hugs}) and the
deeper IRAC imaging provided by the SEDS programme (\cite{seds}),
allowing us to improve the mass estimates for faint ($M_{1400}\sim
-18$) galaxies (green and magenta points in Fig.\ref{masstolight}).

These differences have obvious consequences for the form of any 
GSMF derived from the UV light. Since \cite{gonzalez2011} adopted 
a $M/L$ relations significantly steeper than $\log M\propto -0.4 \times M_{UV}$ 
(they adopted $\log M\propto -0.68 \times M_{UV}$), their inferred 
stellar masses at very faint UV luminosities are
underestimated by an order-of-magnitude with respect to the typical values
derived from a constant $M/L$ ratio relation as found here. 
The resulting GSMFs computed with
the steeper $M/L$ relation are thus inevitably flatter at the faint end 
than the ones derived in the present study.

\begin{figure*}
\includegraphics[width=14cm,angle=-90]{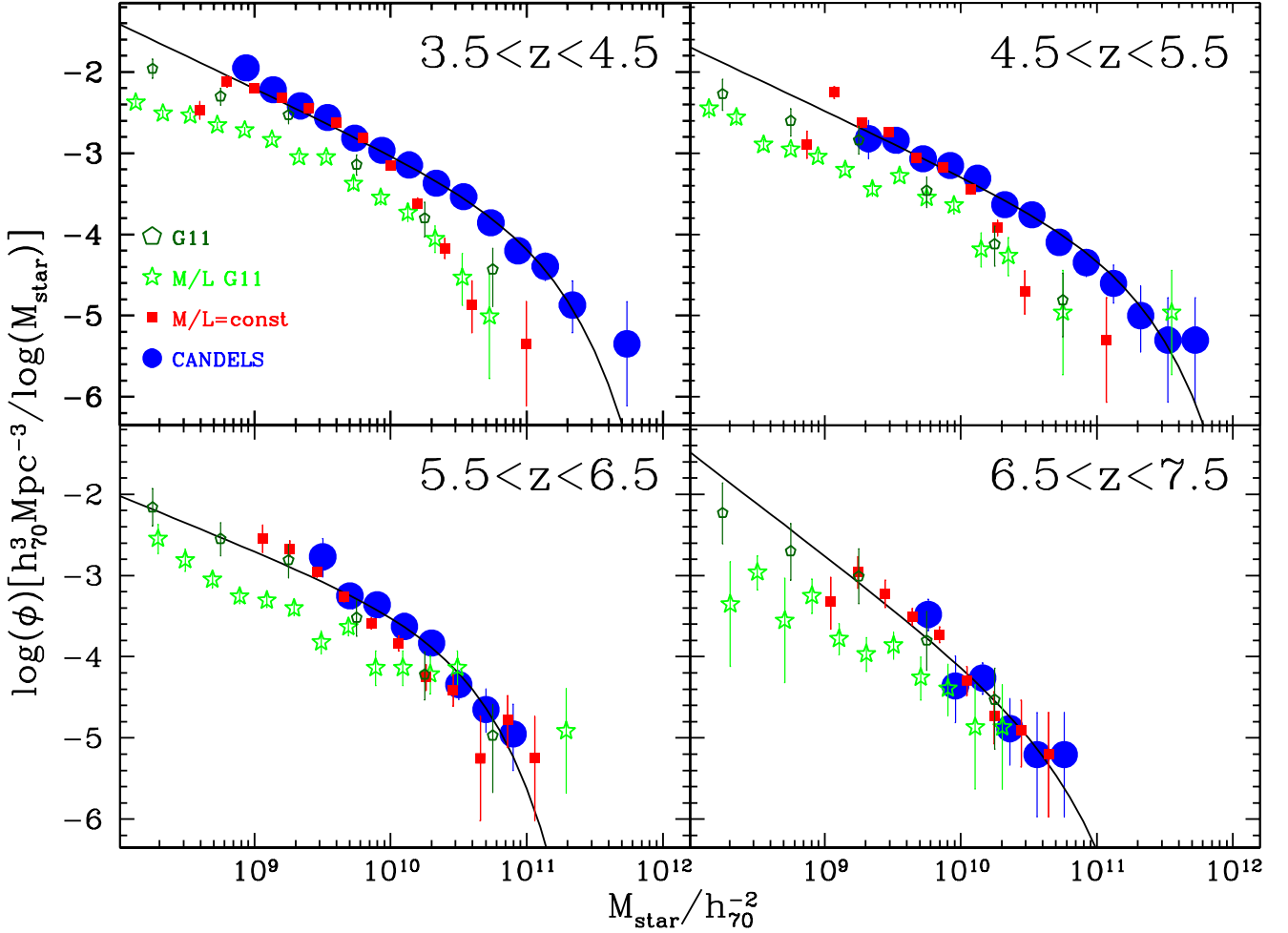}
\caption{The comparison between the mass function derived by assuming a
constant mass-to-light ratio (red squares), the relation between mass and
$L_{1400}$ light derived by \cite{gonzalez2011} (green stars), and the one
derived in this work for the GOODS-South and UDS fields (blue circles).
The error bars show only the Poissonian uncertainties of each point.
The original GSMF presented \cite{gonzalez2011} is shown by dark-green symbols.
}
\label{cmpg11}
\end{figure*}

This effect is clearly shown in Fig.\ref{cmpg11}, where our derived 
GSMFs are compared with those obtained by converting the UV luminosity function
adopting a non-linear functional form for the $M_*/L_{UV}$ ratio. In addition
to a GSMF taken from the literature (\cite{gonzalez2011}, green
pentagons) we show also those obtained from our CANDELS data adopting
either the same $M_*/L_{UV}$ relation as used by \cite{gonzalez2011} (green
starred points) or those obtained adopting our own
$M_*/L_{UV}$ (red dots). We note that the \cite{gonzalez2011} GSMF has
been corrected for incompleteness and for the estimated scatter in the $M/L$ 
relation, and is based on a smaller field, hence its normalization cannot be
immediately compared to that of our GSMF computed using their
$M_*/L_{UV}$ relation.

A few results are immediately evident. First, comparing the GSMF
derived using our $M_*/L_{UV}$ relation with the one derived (from the
same data) using the \cite{gonzalez2011} relation, it is clear that
the latter yields a GSMF that is flatter and appears to extend to
lower masses, since the relation between UV light and mass is steeper
than the one observed in the CANDELS data, as shown in
Fig.\ref{masstolight}. At faint magnitudes, the GSMF derived using our
average $M_*/L_{UV}$ relation agrees very well with the GSMF that we
derive from the full sample. We use this agreement to extend our
fiducial GSMF towards even lower masses, namely to $M=6\times
10^8\,{\rm M_{\odot}}$ at $z=4$ and $M=2\times 10^9\, {\rm M_{\odot}}$
at $z=7$, assuming that losses due to incompleteness are
minimal. These additional points have been marked with blue empty
circles in Fig.\ref{mfallhz}.

Another major discrepancy that emerges from Fig.\ref{cmpg11} concerns
the high-mass end of the GSMF: especially at $z\simeq 4$, the GSMF
derived from our reference sample extends clearly to much higher
masses than all GSMFs computed with some average $M_*/L_{UV}$: our
GSMF (blue circles) extends towards $M\sim 5\times 10^{11}\, {\rm M_{\odot}}$
while the mass functions derived from the UV luminosities are limited
to $M\le 10^{11}\, {\rm M_{\odot}}$. In the next sub-section we will explore the
reasons for this discrepancy.

We note that all these differences tend to disappear at higher
redshifts. Indeed, at $z=6$ and $z=7$, our GSMFs (blue circles in
Fig.\ref{cmpg11}) are consistent with the mass function derived
through the UV luminosity, assuming a constant $M_*/L_{UV}$ ratio (red
squares). There are two possible explanations for this behaviour:
either all the galaxies at high-z have relatively low dust content and
are relatively young or, alternatively, the $H_{160}$-band selection adopted
in this work is missing the more obscured and/or evolved galaxies at
$z>6$. These alternatives can be distinguished by using a deep MIR
selection. Considering the current limitations of IRAC-selected
samples, such as the \cite{caputi11} one, that are significantly
plagued by limited depth and confusion due to poor image resolution,
this issue may not be fully resolved until future {\it JWST} observations.

\subsection{The physical properties of massive galaxies at high redshift}

As shown in Fig.\ref{masstolight}, there are a number of relatively
faint objects ($M_{1400}\sim -18$) which are nevertheless very
massive, with $M\sim 10^{11}\, {\rm M_{\odot}}$ (see also \cite{md14}). While
their absolute number is not very large, it is similar to the number
of UV-bright galaxies of comparable masses, and therefore these objects can 
make an important contribution to the massive end of the GSMF. This is
clearly shown in Fig.\ref{cmpg11}, which illustrates a clear discrepancy,
at $z\le 5.5$,
between the high-mass end of the GSMF derived from the UV-selected
star-forming galaxies (i.e. \cite{stark09,gonzalez2011,lee2012}) and
the one derived in this paper, which has been obtained with a complete
near-infrared sample without the application of colour pre-selection.

\begin{figure}
\centering
\includegraphics[width=9cm,angle=0]{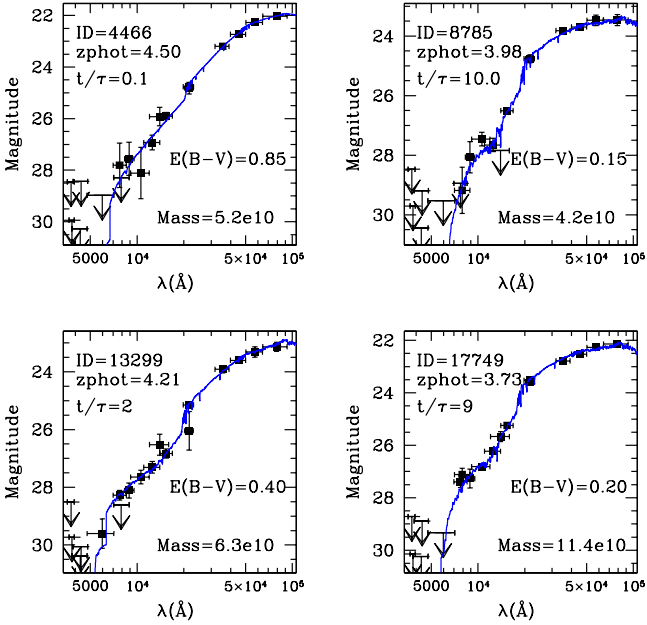}
\caption{Example SEDs of galaxies at $3.5<z<4.5$ with $M>10^{10.5}\,{\rm
M_{\odot}}$ and $M_{1400}>-18.5$. Their SEDs, especially in the IRAC
bands, indicate that they are really massive objects at high redshift
and that the nebular contribution is not dominant. The objects
ID=8785 and 17749 are quite old ($age/\tau\ge 8$), while ID=4466 and
13299 have $E(B-V)\ge 0.4$.  All these galaxies are characterized by
very red colours, and thus cannot be selected by standard LBG
selection criteria based on UV rest-frame colours.  Due to their large
masses, however, they represent a relatively rare population, which
can contribute significantly to the high-mass tail of the mass
function at $z=4$.
}
\label{massivenolbgz4}
\end{figure}

To understand the nature of these galaxies with (relatively) low UV
luminosity but high masses we plot in Fig.\ref{massivenolbgz4} the
spectral energy distributions of the four most massive ($M\ge
10^{10.5}\, {\rm M_{\odot}}$) but faint ($M_{1400}>-18.5$) galaxies in
the GOODS-South field (see also Fig.\ref{masstolight}). The fits to
their observed SEDs indicates clearly that is these objects do indeed
lie at $z>3.5$, but their observed colours are clearly very different
from those used to select LBGs at comparable redshifts. Such objects
are in fact well fitted with dusty star-forming models or
alternatively with passively evolved SEDs. We have checked that they
are not type-1 AGN based on the 4 Ms Chandra observations, but we
cannot exclude that they are X-ray absorbed (Compton thick) AGN
(although they do not show any sign of an unusually steep SED slope in
the IRAC bands). In a separate paper (Merlin et al., in prep.), we
describe in more details the most interesting objects.  Although they
are quite rare, these red galaxies are unambiguously very massive and
represent a major contribution to the high-mass end of the mass
function at $z\simeq 4$.

\section{The evolution of the mass function at high redshift}
\label{GSMFevol}

\begin{figure}
\centering
\includegraphics[width=7cm,angle=-90]{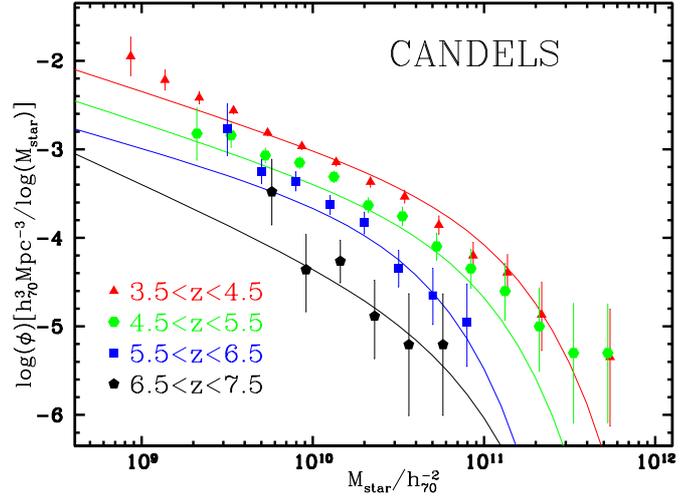}
\caption{The GSMFs from $z=4$ to $z=7$ in the CANDELS UDS and GOODS-South
fields. The error bars take into account the Poissonian statistics,
the Cosmic Variance and the uncertainties derived through the Monte
Carlo simulations. The solid continuous curves show the Schechter
function derived through a best-fit approach which corrects the
observed data points for the Eddington bias.
}
\label{summarymfhz}
\end{figure}

We can finally obtain a full description of the shape and redshift
evolution of the GSMF by fitting a Schechter function to our data.
Before doing so, it is important to carefully consider how the
uncertainties on the estimate of the stellar mass for each galaxy
may impact on the observed shape of the GSMF.

Since the measured masses are randomly perturbed by different noise
sources, the net effect on the observed GSMF is a preferential
transfer of galaxies from the faintest bins toward the more massive
ones, since low mass galaxies are more numerous than brighter
ones. This effect is commonly referred to as ``Eddington bias'',
following \cite{eddington1913}. It is usually believed to affect
mostly the massive side of the GSMF, where the slope is steeper than
at the faint end, but we show below that this does not
necessarily happen in our data set.

For this purpose we use the probability distribution functions
$PDF(M|z)$ (i.e. the probability that a given galaxy in our sample has
a mass $M$ at a redshift $z$) derived from the Monte Carlo simulations
described in Section 4.1.1. We use the individual $PDF(M|z)$s to build
the average $PDF(M|z)$ for each mass bin of the GSMF, at the various
redshifts. While this has already been done in previous analyses
(\cite{ilbert13}) as a function of redshift, the novelty of our
approach is that we explicitly derive the $PDF(M|z)$ as a function of
both redshift and stellar mass. Unsurprisingly, the $PDF(M|z)$s become
wider (implying that masses become less constrained) when redshift
increases and when galaxies become fainter (i.e. less massive).

The procedure adopted is fully described in Appendix B, while we
report here the major results. The first is that the distortions
induced in the shape of the GSMF are progressively larger with
increasing redshift, and become particularly severe at $z\simeq
7$. This is due to the fact that, as naively expected, the intrinsic
errors on the estimate of individual galaxy stellar masses 
increase with increasing redshift (Fig.\ref{sumpm}).

The second effect is that, contrary to what was found in previous
analyses (e.g. \cite{ilbert13}), especially at $z=4$, the accuracy at
the bright end is here good enough to keep the overall shape essentially
unaffected. On the contrary, the increase in the noise at the faint
end (that was never taken into account in previous estimates) induces
a non-negligible steepening of the observed faint end. Both effects
are clearly visible in Fig.\ref{convphiz4ilb}. At $z\ge 6$, instead,
the large errors in mass can affect also the shape of the GSMF at the
massive side (see Fig.\ref{convphiz6}).
It's worth noting that the Eddington bias corrections described here
depend critically on the shape of the adopted PDFs in mass. As a
cautionary note, we warn the reader that the unambiguous estimation of
the best fitting GSMF is subject to the correct derivation of the
PDF(M).

We include these effects in our fit of the GSMF. We adopt for the
GSMF a standard parametrization with a single Schechter function with
free parameters $\alpha$, $M^*$ and $\Phi^*$. We have performed a
best-fit on the data points obtained with the $V/V_{max}$
approach. Although this is statistically less rigorous than other
approaches (like the STY method), it has two advantages in our case.
The first is that we can fit both the GSMF and its extension to 
lower masses which we derived
by assuming a constant $M/L$ ratio (as described in the previous
section). The second advantage is that we can correct the observed data
points for the Eddington bias. The
fit is performed as follows: for any possible combination of the
Schechter parameters $\alpha$, $M^*$, and $\Phi^*$, we compute the
convolved GSMF using the estimated average $PDF(M|z)$s and we compare
it with the observed mass function. We scan the three-dimensional
space of the free parameters of the Schechter function to find the
best fit solution by a $\chi^2$ minimization. The error bars of the GSMF
considered here include both the Poissonian error bars from the
$1/V_{max}$ procedure as well as the uncertainties due to the other
effects (photometric redshifts, photometric scatter) that are
estimated with our Monte Carlo simulations described before. We
include also the errors due to cosmic variance.

Fig.\ref{summarymfhz} shows the resulting GSMFs from $z=4$ to $z=7$ in the
combined UDS and GOODS-South fields. Our GSMFs extend toward low
masses: $10^9\, {\rm M_{\odot}}$ at $z=4$ and $6\times 10^9\, {\rm M_{\odot}}$ at 
$z=7$ (even lower, converting UV light to stellar masses). Since we
explicitly correct for the Eddington bias, the best-fit GSMFs do not
follow the raw binned data; as mentioned above, the recovered slope is
less steep in all cases, and the massive side at high redshift is
shifted toward lower masses, though the latter trend is not
particularly significant.
We note also that in Appendix C we
provide for comparison the results of the fitting procedure neglecting
any effect from the Eddington bias.

\begin{figure}
\includegraphics[width=9cm,angle=0]{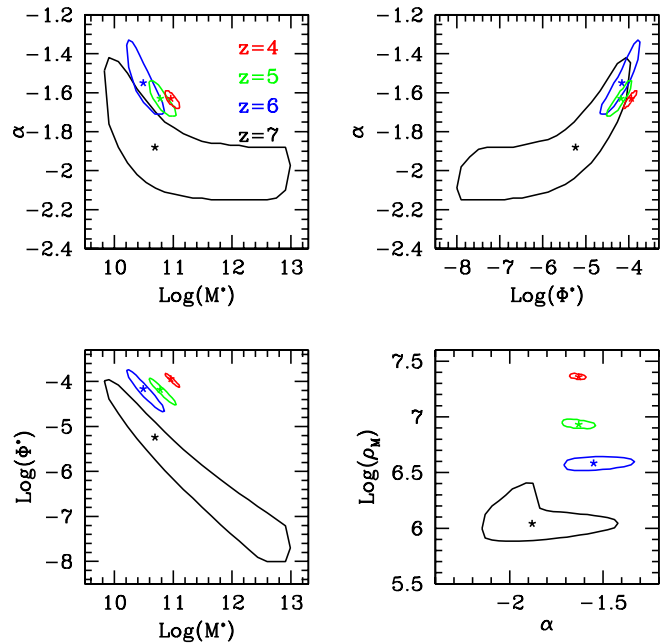}
\caption{The evolution of the three parameters ($\alpha$, $M^*$, $\Phi^*$)
of the GSMF with redshift. The bottom-right panel shows the dependencies of
the Stellar Mass Density ($\rho_M$ in unit of $M_{\odot}Mpc^{-3}$) from
the parameter $\alpha$. The stars mark the position of the best fit
of the observed GSMF with a Schechter function, obtained by correcting for the
Eddington bias as explained in detail in the Appendix B.
A clear trend with redshift is evident up to $z=6$, while at $z=7$ the
parameters are basically unconstrained.
}
\label{alpha}
\end{figure}

Table \ref{tabgsmf} summarises the $\chi^2$ best fit values of the
free parameter of the GSMF, again corrected for the Eddington bias.
We have also been able to compute the errors on such parameters, at
the 1 $\sigma$ confidence level, by scanning the three dimensional
volume of the three parameters. The same results are also shown in
Fig.\ref{alpha}, that reports the evolution of $\alpha$, $M^*$ and
$\Phi^*$ with redshift, along with their uncertainties.

We find that the best-fit parameters are reasonably well constrained
up to $z=6$, such that meaningful conclusions on the evolution of the
various parameters can be derived. In the last redshift bin, the
uncertainties (that are due to both the intrinsic errors on the
stellar masses as well as to the small size of the sample) are much
larger, and results are only tentative.

We find that the slope of the low-mass side of the GSMF is almost
constant, with $\alpha \simeq -1.6$ from z=4 to z=6, showing no
evidence for a steepening with redshifts. This slope however
is significantly steeper than the GSMF slope up to $z\simeq 1$, indicating
that a progressive steepening must occur in between. \cite{santini12}
presented the first evidence for such a trend, but this will 
clearly need to be verified with much deeper and wider surveys like CANDELS.

Most of the evolution of the GSMF appears to be due to a combination
of density evolution ($\Phi^*$ grows at lower-z) and mass evolution
($M^*$ increases with cosmic time).

In the last redshift bin, at $z=7$, the large errors reflect strong
degeneracies between the Schechter parameters, and the best
fit values are much less constrained, so that any conclusion is
preliminary. The only robust result is a further decline in the total
stellar mass density $\rho_M$, that is steadily decreasing as redshift
increases, at several $\sigma$ of significance (Fig.\ref{alpha},
bottom right panel).

\begin{table}
\caption{Mass Function best fit parameters}
\label{tabgsmf}
{\centering
\begin{tabular}{c | c c c r}
\hline\hline
Redshift & $\alpha$ & $\log(M^*/{\rm M_{\odot}})$ & $\log(\Phi^*)$ & $N_{gal}$ \\
\hline
$3.5<z<4.5$ & $-$1.63$\pm$0.05 & 10.96$\pm$0.13 & $-$3.94$\pm$0.16 & 1293 \\
$4.5<z<5.5$ & $-$1.63$\pm$0.09 & 10.78$\pm$0.23 & $-$4.18$\pm$0.29 &  370 \\
$5.5<z<6.5$ & $-$1.55$\pm$0.19 & 10.49$\pm$0.32 & $-$4.16$\pm$0.47 &  126 \\
$6.5<z<7.5$ & $-$1.88$\pm$0.36 & 10.69$\pm$1.58 & $-$5.24$\pm$2.02 &   20 \\
\hline
\end{tabular}
}
The parameters $\alpha$, $\log(M^*)$, and $\log(\Phi^*)$ of the Schechter
function at $z=4-7$ derived through $\chi^2$ fitting to the observed
data points shown in Fig.\ref{mfallhz}, after correcting for Eddington bias. 
Uncertainties refer to to 1$\sigma$ confidence intervals.
$N_{gal}$ is the number of galaxies (GOODS-South + UDS fields)
in each redshift bin used to compute the GSMF. 
\end{table}

Overall, the mild evolution in both $\log(M^*)$ and $\log(\Phi^*)$,
together with an almost constant $\alpha$,
seems to indicate that the mass assembly rate was somewhat similar for
massive galaxies and for the less massive ones. This behaviour can be
linked to the apparently slow (or negligible) 
redshift evolution of the apparent SF quenching mass as has been 
found at lower redshifts in the
COSMOS data by \cite{peng10,peng12} and \cite{ilbert13}, although
clearly the physical processes ongoing in high-redshift galaxies may
be very different.

\section{The stellar mass density at $3.5\le z\le 7.5$}

\begin{figure}
\includegraphics[width=9cm,angle=0]{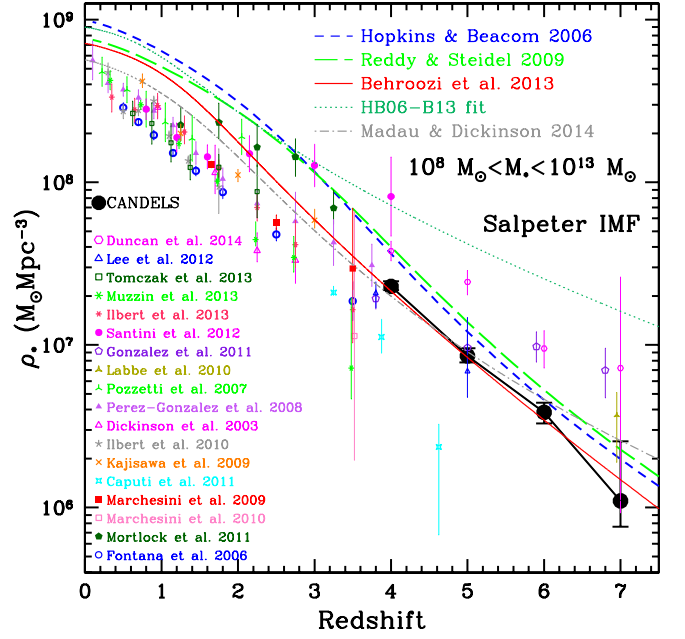}
\caption{The redshift evolution of the stellar mass density (SMD) at
$3.5<z<7.5$ obtained by integrating the GSMFs presented in this paper
(black points). The evolving SMD at high redshift is compared to the
lower redshift results from different surveys. $\rho_M$ is in units of
${\rm M_{\odot}Mpc^{-3}}$ and has been obtained by integrating the
best-fitting mass functions from $M_{min}=10^8\,{\rm M_{\odot}}$ to
$M_{max}=10^{13}\, {\rm M_{\odot}}$.
All the SMDs have been converted to a Salpeter IMF for comparison.
The error bars of the CANDELS data have been computed
using the same Monte Carlo simulations developed to derive the
uncertainties on the Schechter function parameters. The short-dashed
line is the stellar mass density obtained integrating over cosmic time
the star formation rate density (SFRD) of \cite{hb06}. The
long-dashed line is the SMD inferred from the SFRD of \cite{rs09}. The
solid line is the SMD obtained from the SFRD of \cite{behroozi13},
while the dotted line is the SMD derived by the new fit of the
\cite{hb06} carried out by \cite{behroozi13}. The dotted-dashed line
shows the SMD derived from the SFRD of \cite{md14}. All the stellar
mass densities obtained by integrating the different SFRDs assume a
constant recycling fraction of 28\%.
}
\label{smd}
\end{figure}

Starting from the best fit of the GSMF and from its uncertainties, we
derive the stellar mass density (SMD) by integrating the best-fit GSMF
from $M=10^{8}\,{\rm M_{\odot}}$ to $M=10^{13}\,{\rm M_{\odot}}$. These
limits have been chosen to facilitate the comparison with previous
SMD estimate in the literature. To compute the
associated error, we consider all the possible combinations of the
parameters $\alpha$, $M^*$, and $\Phi^*$ which are still compatible
with the observed GSMFs at 68\% c.l. and for each of them we compute
the integral in mass, thus finding the minimum and maximum range of
the SMD at the 1-$\sigma$ level. The results are shown in
Fig.\ref{smd}, that shows the SMD, $log(\rho_M)$, at $3.5<z<7.5$
obtained in the CANDELS-UDS and GOODS-South fields (black points).
Table \ref{tabsmd} summarizes the values of the SMD
at $3.5<z<7.5$ with its uncertainties.

In Fig.\ref{smd} we also compare our evolving SMD with those derived
by different surveys in various redshift ranges. We translate all the
SMDs to the Salpeter one adopted in this paper. At intermediate
redshift ($2<z<4$) the scatter is of the order of 0.3-0.5 dex and can
be due to cosmic variance or photometric redshift uncertainties.  At
low redshift ($z<2$) the scatter is reduced, and it is probably driven
by statistical uncertainties. At $z>4$ the different SMDs are in
agreement with each other, with a few exceptions.

The redshift evolution of the GSMF derived above
results in a rapidly evolving SMD at early cosmic epochs.  Our
results are in agreement with those of \cite{duncan14}, which are based
on a similar analysis of earlier CANDELS data, and indicate a faster
evolution than those of \cite{gonzalez2011}, whose SMD is instead
evolving slowly with redshifts, probably due to the different method
adopted (see Section 5.2).

\begin{table}
\caption{Stellar Mass Density at $3.5<z<7.5$}
\label{tabsmd}
\centering
\begin{tabular}{c | c c c}
\hline\hline
Redshift & $log(\rho_M)$ & Min $log(\rho_M)$ & Max $log(\rho_M)$ \\
\hline
$3.5<z<4.5$ & 7.36 & 7.34 & 7.39 \\
$4.5<z<5.5$ & 6.93 & 6.89 & 6.98 \\
$5.5<z<6.5$ & 6.59 & 6.52 & 6.64 \\
$6.5<z<7.5$ & 6.04 & 5.88 & 6.41 \\
\hline
\end{tabular}
\\
The stellar mass density (SMD) $log(\rho_M)$ is derived from the best
fit of the GSMF, integrating it from $M=10^{8}\,{\rm M_{\odot}}$ to
$M=10^{13}\,{\rm M_{\odot}}$ and assuming a Salpeter IMF. The best fit
of the GSMF takes into account the Eddington bias, as discussed in the
text and in Appendix B. The SMD $\rho_M$ is in units of ${\rm
M_{\odot}Mpc^{-3}}$. The minimum and maximum SMDs indicate the
1-$\sigma$ range (i.e. 68\% confidence interval).
\end{table}

In Fig.\ref{smd} we compare the SMD evolution in redshift with the
integrated value of the star formation rate density (SFRD) over the
cosmic time. We assume a Salpeter IMF and a constant gas recycling
fraction of 28\% (\cite{nagamine06,santini12,md14}). Different
renditions of SMD can be obtained depending on the assumed scenario for the
global SFRD. The short-dashed line is the SMD obtained using the SFRD
of \cite{hb06}. The long-dashed line is the analogous curve for the
SFRD by \cite{rs09}. The solid line is the SMD obtained by the SFRD of
\cite{behroozi13}, while the dotted line is the SMD derived by the new
fit of the \cite{hb06} data points carried out by
\cite{behroozi13}. The dot-dashed line shows the SMD obtained
integrating the SFRD of \cite{md14}. The SFRDs of \cite{hb06} and
\cite{rs09} show a peak around $z\simeq 2$, while the one of
\cite{behroozi13} and their new fit to the data points of \cite{hb06}
have a maximum at lower redshift ($z\sim 1$), similar to the
\cite{md14} SFRD. The latter has a slightly lower normalization at
$z\sim 1$ than the other parametrizations and shows a milder
evolution towards high redshifts.
The typical errors on the SFRD determination are $\sim 0.1$ dex at $z\le 0.9$
and $\sim 0.2-0.3$ dex at $z\ge 1.7$ (see Table 7 of \cite{behroozi13}).

At face value, the results shown in Fig.\ref{smd} indicate that
the growth in SMD derived here at high redshifts ($z>4$) is in 
agreement with that inferred from integrating the SFRD histories 
presented \cite{behroozi13} or \cite{md14}, and
lower than the one derived in
the new fit of the \cite{hb06} dataset by \cite{behroozi13}. Indeed,
our SMD can be reproduced by a SFRD which is evolving at high
redshifts at the same rate of the parametrizations of \cite{hb06},
\cite{rs09}, \cite{behroozi13} or \cite{md14}. In addition, the
normalization of our measurements is slightly lower than the one of
\cite{hb06} and \cite{rs09}. This indicates that at $z\ge 4$ the SMD
and SFRD are in very good agreement. We must stress however that this
is true when the GSMF and the SFRD are integrating till low level of
stellar masses and star formation rate - if we restrict the comparison
to specific ranges of stellar masses (or, more accurately, of parent
halo mass) the results could well be discrepant.

Of course, our analysis cannot say anything new about the puzzling
disagreement between the growth of SMD and the time integral of SFRD
at $z<2$, as noticed also by \cite{santini12} and \cite{md14}. In this
redshift range we expect that both the SFRD and the SMD estimates are
robust and not affected by incompleteness.

The exact size of this discrepancy, and whether it represents a major
problem is still matter of considerable debate. Many explanations have
been proposed, including an underestimate of the mass density in local
galaxies (\cite{bernardi13}), an effect of ``outshining'' due to
recent stellar populations that lead to an underestimate in stellar
masses (\cite{maraston10,massreview}), the scattering of stars in the
intracluster light (ICL) during galaxy mergers in group or clusters,
or even time-varying IMFs. At face value, the agreement that we find
at $z>4$ is apparently in contrast with a strong evolution of the IMF
at high redshift, especially at $z>6$ as advocated for example by
\cite{chary08}. They argued that the IMF at $z\ge 6$ could be
different from the Salpeter one in order to have the Universe
reionized by these redshifts and to obtain an agreement with the WMAP
measurement of the optical depth $\tau_{es}$
(\cite{spergel07}). Clearly, we need more accurate estimates of both
the SFRD and SMD to definitely address this issue.


\section{Summary and Conclusions}

We have combined wide and deep {\it HST} {\it Spitzer} and VLT 
observations in the CANDELS UDS (\cite{galametz}), GOODS-South (\cite{guo}),
and HUDF (\cite{udf,hudf09}) fields to study the evolution of the
Galaxy Stellar Mass Function in four redshift bins between $3.5\le
z\le 7.5$. The {\it HST} data cover 369\,arcmin$^2$ down to a magnitude
limit (at 5$\sigma$ in apertures of 2 times the FWHM) of
$H_{160}=26.7$ and 27.5 for the UDS and GOODS-South field,
respectively, reaching a depth of $H_{160}=28.5$ in a limited area of
5\,arcmin$^2$ covered by the HUDF region.

In addition to the imaging data already adopted by \cite{galametz} and
\cite{guo}, we have included the deep $K$-band images obtained from
the VLT Hawk-I survey HUGS (\cite{hugs}), reaching $K \simeq 26.5$ at
5-$\sigma$ over the GOODS-South and HUDF fields. The deep IRAC images
from the SEDS survey (\cite{seds}) are also a crucial ingredient of
our data base. Finally, we have also added deep $B$-band imaging from
VIMOS at VLT (\cite{nonino}). The photometric technique adopted to
de-confuse the ground-based and {\it Spitzer} images at the faintest
limits is described in \cite{galametz} and \cite{guo}.

The high-quality photometry from the near-UV to 8 $\mu$m, has been
used to derive accurate estimates of photometric redshifts and stellar
masses. For the photometric redshifts we have adopted the innovative
technique explored in \cite{dahlen13}, which combines
different photometric redshift solutions with a Bayesian technique to 
provide probability distribution functions in redshift, taking into account
the biases and the scatter of each individual solution. This approach
allows us to obtain photometric redshifts that are significantly more
accurate than each individual method, as demonstrated by the
comparison with a sample of $>$2500 galaxies with robust spectroscopic
redshifts. We find an absolute scatter of $\sim$0.03 and an outlier
fraction of 3.4\%. At $3.5\le z\le 7.5$ the photometric redshifts
have similar precision, with an absolute scatter of 0.037 and 7\% 
outliers, based on a sample of 152 robust spectroscopic redshifts.

With this exquisite data set at hand, we have explored different
recipes to derive the stellar masses, focusing on the redshift range
$3.5<z<7.5$. Following a standard approach, we have used the BC03
spectral synthesis code to predict galaxy colours for a wide range of
galaxy properties, including different star-formation histories,
ages, metallicities and dust content, and derived galaxy stellar mass from the
best-fitting spectral template at the photometric redshift. With
respect to previous studies, we have significantly increased the
breadth of the parameter space spanned by the models. First, we have
tested different parametrizations of the star-formation histories
allowed for galaxies: in addition to the standard exponentially-declining
models, we have allowed for ``Inverted-$\tau$'' models
(SFH$\propto exp(+t/\tau)$) as well as for ``Delayed'' star-formation
histories (SFH$\propto t^2/\tau \times exp(-t/\tau)$). For all these
models, we have also tested how stellar masses change when full
nebular emission (both lines and continuum) is included using the
prescription of \cite{schaerer2009}.

We have carefully explored the impact of these different assumptions on the
derived GSMF, as well as of other systematics, such as cosmic variance
(field-to-field variation) and the contribution of AGN
(selected via X-ray emission, spectral identification or variability).
We show that, quite reassuringly, the stellar masses and the derived
stellar mass functions turn out to be quite stable against different
choices of the adopted SFH. The inclusion of the contribution of
nebular lines and continuum in the SED of the galaxies is also not
dramatic, as it systematically lowers the stellar mass estimates by
0.05-0.20 dex at $z\ge 4$.

We have found that photo-$z$ errors
are the largest sources of uncertainty in the derived GSMF, even
larger than cosmic variance. In particular, we have shown that the
adoption of different recipes for the computation of photometric
redshift (on the same photometric sample) can be the largest source of
uncertainty - an important lesson to keep in mind when one compares
results from different surveys, where different recipes for photo-$z$
are used. These two error terms increase towards large redshift, from
0.1 dex at $z=4$ to slightly less than 0.3 dex at $z=7$.

We then  used our sample of 3307 galaxies at $3.5<z<7.5$ selected via
photometric redshifts (or spectroscopic when available) to compute the
GSMF of galaxies down to low mass limits.
For consistency with previous works, we
have adopted as reference masses those derived adopting only
exponential declining star formation histories, without the
contribution of nebular lines nor continuum. At $z=4$ we reach a
completeness mass limit of $M=10^9\, {\rm M_\odot}$, which increases 
progressively with increasing redshift to 
$M=6\times 10^9\, {\rm M_\odot}$ at $z=7$.

A crucial ingredient in our analysis is a careful estimate of how the
uncertainties on the measurement of the galaxy stellar mass affect the
derived GSMF - the so called Eddington bias. For this purpose we use
the probability distribution functions $PDF(M|z)$s (i.e. the
probability that a given galaxy in our sample has a mass $M$ at a
redshift $z$) derived from the Monte Carlo simulations as described in
Section 4.1.1. At variance with previous analyses (\cite{ilbert13}),
our approach is novel in that we explicitly let the $PDF(M|z)$ vary as
a function of redshift and stellar mass. Unsurprisingly, the
$PDF(M|z)$s become wider (implying that masses become less
constrained) when redshift increases and when galaxies become
fainter. When these effects are properly taken into account, we show
that Eddington bias not only flattens and boosts the apparent GSMF at
high masses (albeit much less severely than in other surveys,
especially at $z=4$, due to the excellent quality of our data) but
also induces an apparent steepening of the GSMF at the faint side.

A first main result of our analysis is the evidence that, at least at
$z\simeq 4$, the massive side of the GSMF is dominated by galaxies
that are not ordinary Lyman Break Galaxies. We have shown that there
exists a significant population of $z_{phot}\simeq 4$ galaxies that are
intrinsically redder than LBGs, i.e. that are faint in the UV
rest-frame but are bright in the IR bands, indicating that they are
either highly-obscured by dust or old/evolved objects, and thus they
have large stellar masses. While the existence of passively evolved
galaxies at $z=4$ has also been shown recently by \cite{straatman13}
using the ZFOURGE dataset and deep {\it Herschel} observations, in agreement
with early results by \cite{fontana09} in the GOODS-South field, we
show here that they do contribute significantly 
to the high-mass end of the GSMF, certainly at $z \simeq 4$.

We were led to this evidence by looking at the difference between our
GSMF and the previous ones that were computed from $L_{UV}$ luminosity
functions, scaled to mass functions by adopting average relations
for the $M_*/L_{1400}$ ratio. The biggest difference is found with respect to
the GSMF derived by \cite{gonzalez2011} at $z \simeq 4$. 
This is clearly shown by a comparison
between our GSMF and the one that we derive by assuming that all
galaxies have a constant $M_*/L_{1400}$. In our sample, that is mostly
composed of star-forming galaxies with modest dust obscuration,
we find that the relation between $M_*$ and
$M_{1400}$ can be fitted with a linear slope $-0.4$, albeit with large
scatter, that is equivalent to a constant $M_*/L_{1400}$. This result
is in agreement with recent results on smaller CANDELS data sets
(\cite{salmon13} and \cite{duncan14}), and is at variance with
\cite{gonzalez2011}, who found a decreasing trend of the
$M_*/L_{1400}$ with decreasing luminosity. We ascribe this difference
to the superior quality of our data, that at $z\simeq 4$ benefits from 
the ultradeep HUGS data in GOODS-South for measuring 
the rest-frame optical luminosity of even the faintest galaxies.

At higher redshifts ($z=6-7$) the GSMF inferred by \cite{gonzalez2011}
is in better agreement with our estimates. At first glance this
agreement can indicate that, at very high redshift, the majority of
the galaxy population comprises star-forming galaxies with negligible
dust extinction and that the population of dusty and/or evolved
galaxies has virtually disappeared at very early cosmological
epochs. However, an alternative explanation could be that we are not
sensitive to those red galaxies at extreme redshift, since our study
is based on $H$-band selected catalogues, which at $z=7$ are sampling
purely the far UV ($\lambda\sim 2000\AA$) rest-frame emission from
young stars. Thus the disappearance of dusty or old galaxies could
simply be a selection effect. In the future, deep infrared selected
samples at longer wavelengths will be fundamental to answering this
question.

The central result of our paper is related to the evolution of the
GSMF as a function of redshift, in which we detect a clear decrease with
increasing redshift, at both high and low masses. Adopting a
Schechter parametrization at all redshift, and including the effects
due to the Eddington bias, we quantify this evolution by looking at
the best fit parameters $\alpha$, $M^*$ and $\Phi^*$ along with their
errors. We find that these parameters are well constrained up to
$z=6$, while at $z=7$ the increasing uncertainty in the estimate of
stellar masses and the small size of our sample prevent us from
deriving robust constraints on these parameters, and only the total
mass density $\rho_M$ is robustly derived. We find that the slope of
the low mass side of the GSMF is relatively constant at about
$\alpha=-1.6$ at $4<z<6$. This slope is, however, significantly steeper
than the slope up to $z\simeq 1$, indicating that a progressive
steepening must occur at intermediate redshifts. We also find that
most of the evolution of the GSMF results from a combination of
density evolution and mass evolution (both $\Phi^*$ and $M^*$
increase with cosmic time). In particular, $M_*$ shifts from
$log(M_*/{\rm M_\odot})=11$ (a value close to $M_*$ at lower redshift) to
$log(M_*/{\rm M_\odot})\simeq 10.4$ at $z=6$. We caution however that both
trends are only marginally significant.

Adopting our parametrization of the GSMF, we computed the stellar mass
density (SMD) of galaxies at $z=4-7$ by integrating the observed GSMF
down to $10^8 {\rm M_\odot}$ and compared it with the integrated value of
the star formation rate density (SFRD) over cosmic time, assuming a
constant recycling fraction of 28\%. We found that the time integral
of the SFRD is in overall agreement with our measurements at $z>4$
when we adopt the estimated parametrization by \cite{behroozi13} or
\cite{md14}, although other parametrizations can be quite discrepant.

The observed evolution of the GSMF with redshift can also provide 
indications on the expected growth-rate of galaxies at the high-mass
end. Using a simple argument based on object number conservation,
without taking into account the correction for merging as done in
\cite{papovich2011} and in \cite{behroozi13}, from
Fig.\ref{summarymfhz} we can derive for galaxies at a fixed density of
$10^{-5}h_{70}^3 Mpc^{-3}/log({\rm M_\odot})$ a mean SFR of $\sim 290\,{\rm 
M_{\odot} yr^{-1}}$ from $z=6$ to $z=4$. To compute this quantity, we have
adopted the best fit of the GSMFs with the Eddington bias correction,
as detailed in Section 6.

If we compare this star formation rate with the SFR Function of
\cite{smit12} at $4\le z\le 6$, it is clear that the massive galaxies
we have found at the massive tail of the GSMF at $z\ge 4$ are probably
not represented by the typical UV bright (and characterized by low
dust extinction) population found by {\it HST}, but represent
plausibly a dusty starburst phase in the life of these galaxies.
Thus, the massive galaxies we find in the exponential tail of the GSMF
at $z\ge 4$ can plausibly be the results of very active phases, which
have been detected in the sub-mm regime and that will represent an
exploratory field for ALMA in the near future. Alternatively, the SFR
derived by UV light using LBGs at high-z can be underestimated by a
factor of 2-3, as recently suggested by \cite{castellano14}. An
important aspect here is related to the intrinsic uncertainties of the
simple number conservation approach adopted, which does not take into
account the effects of merging and the one-to-one correspondence
between mass and UV light/SFR. A detail comparison with theoretical
predictions can partially alleviate this issue.

We have demonstrated the unique synergy of CANDELS, HUGS, and SEDS in
the analysis of the GSMF at high redshifts extending to low stellar
masses. In particular these surveys take advantage of a combination of
wide and deep data, which allows us to reduce the impact of cosmic
variance and the degeneracies between the parameters of the Schechter
function, used to fit the observed GSMF. The final CANDELS survey
will more than double the present area available, adding data from the
COSMOS, EGS, and GOODS-North fields. It will be complemented by the
deep HUDF12 data and also by {\it HST} images of the Hubble Frontier
Fields initiative, offering a further a step forward in our
understanding of the high-redshift GSMF before the advent of the {\it
JWST} and the ELTs.


\begin{acknowledgements}
We warmly thank the referee for her/his constructive report.
We acknowledge financial contribution from the agreement ASI-INAF I/009/10/0.
This work is based on observations taken by the CANDELS Multi-Cycle
Treasury Program with the NASA/ESA HST, which is operated by the
Association of Universities for Research in Astronomy, Inc., under
NASA contract NAS5-26555.
Observations were also carried out using the Very Large Telescope at
the ESO Paranal Observatory under Programme IDs LP186.A-0898, LP181.A-0717,
LP168.A-0485, ID 170.A-0788, ID 181.A-0485, ID 283.A-5052 and the ESO Science
Archive under Programme IDs 60.A-9284, 67.A-0249, 71.A-0584, 73.A-0564,
68.A-0563,
69.A-0539, 70.A-0048, 64.O-0643, 66.A-0572, 68.A-0544, 164.O-0561,
163.N-0210, 85.A-0961 and 60.A-9120.
This work is based in part on observations made with the Spitzer Space
Telescope, which is operated by the Jet Propulsion Laboratory, California
Institute of Technology under a contract with NASA. Support for this work
was provided by NASA through an award issued by JPL/Caltech.
AF and JSD acknowledge
the contribution of the EC FP7 SPACE project ASTRODEEP (Ref.No:
312725). JSD also acknowledges the support of the Royal Society via a Wolfson
Research Merit Award, and the support of the ERC through an Advanced Grant.
\end{acknowledgements}

\Online

\begin{appendix}

\section{Comparison of the GSMFs obtained with different recipes}

\begin{figure}
\includegraphics[width=7cm,angle=-90]{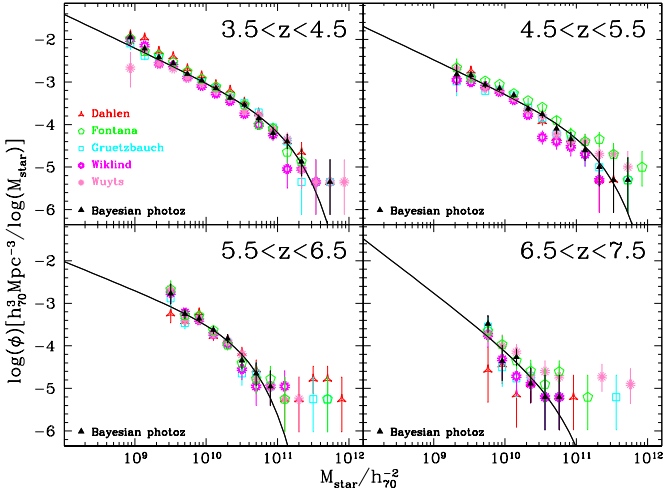}
\caption{Comparison of the GSMF obtained with different photometric redshift
recipes. The stellar mass function of galaxies at $3.5\le z\le 7.5$
in the CANDELS GOODS-South and UDS fields with the Bayesian
photometric redshifts is shown by the black triangles and the solid
continuous curves. The red triangles, green circles, cyan triangles,
magenta and pink asterisks show the GSMFs obtained using the
individual photometric redshifts of five different groups that have
been used to derive the Bayesian photo-z described in \cite{dahlen13}.
}
\label{rec}
\end{figure}

\begin{figure}
\includegraphics[width=7cm,angle=-90]{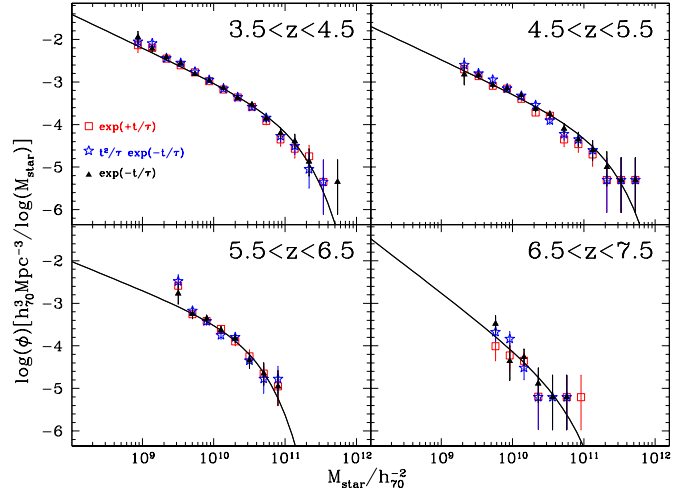}
\caption{Comparison of the GSMF obtained with different star-formation
histories. The stellar mass function of galaxies at $z\ge 3.5$ in the
CANDELS GOODS-South and UDS fields with our standard BC03 fit
(exponentially-declining SFHs) is shown by the black triangles and the
solid continuous curves. The red squares and the blue stars show the
GSMFs derived using exponentially-increasing and a $t^2/\tau \times
exp(-t/\tau)$ SFH, respectively. All these star-formation histories
have been tested without the nebular contribution.
}
\label{sfh}
\end{figure}

\begin{figure}
\includegraphics[width=7cm,angle=-90]{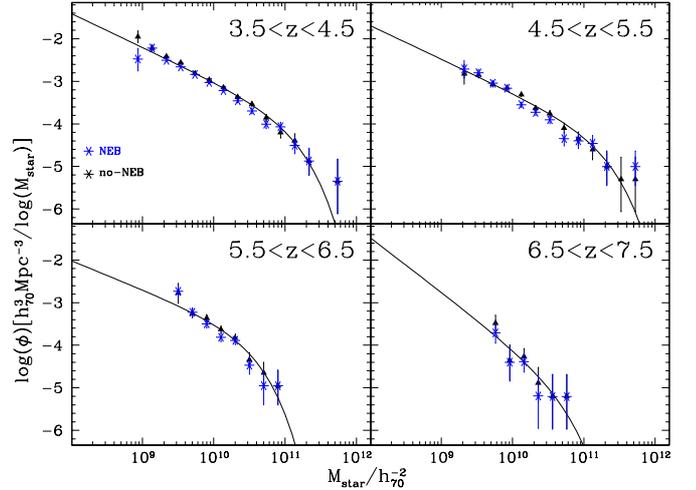}
\caption{Comparison of the GSMF with and without an allowed nebular
contribution. The stellar mass function of galaxies at $3.5\le z\le
7.5$ in the CANDELS GOODS-South and UDS fields with our standard BC03
fit (exponentially-declining SFHs and no nebular contribution) is
shown by the black triangles and the solid continuous curves. The blue
asterisks show the GSMF derived using BC03 models and
exponentially-declining SFHs, but this time including the contribution
of nebular lines and nebular continuum.}
\label{nebmf}
\end{figure}

\begin{figure}
\includegraphics[width=7cm,angle=-90]{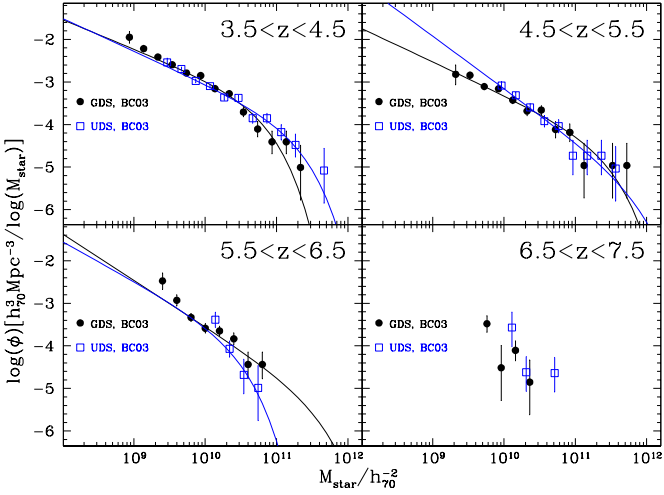}
\caption{The stellar mass function of galaxies at $3.5\le z\le 7.5$ in the
CANDELS UDS field (blue squares) is compared with the one derived from
the GOODS-South data (black circles). At $6.5<z<7.5$ the fit to the
data point was not derived due to the small range in mass of the
observed data in the individual fields.
}
\label{udsgds}
\end{figure}

\begin{figure}
\includegraphics[width=7cm,angle=-90]{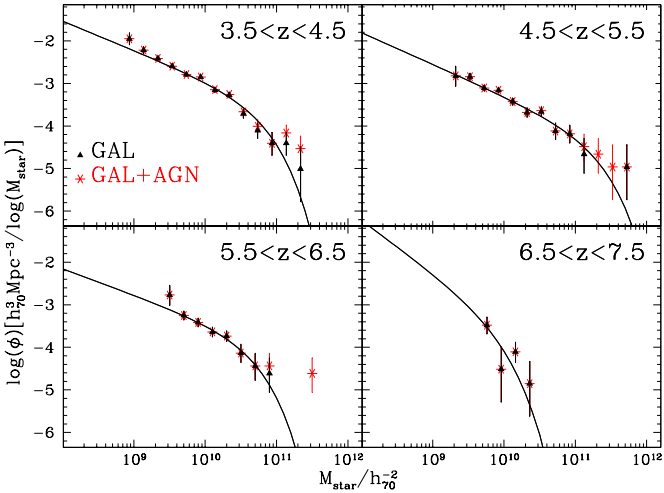}
\caption{Comparison of the GSMF with and without AGN.
The stellar mass function of galaxies at $3.5\le z\le 7.5$ in the CANDELS
GOODS-South field considering only normal galaxies is shown by
the black triangles. The red asterisks show
the derived GSMF including AGN.
}
\label{cmpagn}
\end{figure}

In this section we show the comparison of the GSMF derived with our
standard approach (GOODS-South and UDS fields, Bayesian photometric
redshifts, BC03 library (\cite{bc03}), exponential declining SFHs, no
nebular contribution, no AGN) against the different GSMFs obtained
varying only one ingredient at a time. Fig.\ref{rec} shows the GSMF
at $3.5<z<7.5$ obtained with different photometric redshift recipes
available within the CANDELS team (\cite{dahlen13}). Fig.\ref{sfh}
investigates the impact of different SFHs, Fig.\ref{nebmf} shows the
effect of the nebular contribution on the mass estimates. Finally,
Fig.\ref{udsgds} and \ref{cmpagn} illustrate the cosmic variance
effect (or field-to-field variation) and the contribution of AGN to
the GSMF. In this case we must point out that the information
available for the UDS field (X-ray coverage, deep spectroscopy,
variability) is not equivalent to the rich dataset in GOODS-South, so
at the present stage only some type-1 AGN can be found in the current
UDS galaxy sample adopted in this paper. For this reason only the
GOODS-South field has been used to carry out the comparison in
Fig.\ref{cmpagn}.

\section{The correction of the Eddington bias}

As already discussed in the main text, the stellar mass of a single
galaxy is not unequivocally determined due to the uncertainties on the
photometric redshift determination and on the mass-to-light conversion
adopted to derive the stellar mass from the observed SED. These
uncertainties must be properly taken into account when the observed
data points of the mass function are fitted with a parametric
function. This is the so-called Eddington bias (Eddington 1913), and
as we will show here it could systematically affect the derivation of
the Schechter parameters for the observed mass function.

To correct for this effect we used the probability distribution
function in mass $PDF(M|z)$ that we derived for each individual galaxy
during our SED fitting
procedure, as described in Section 4.1.1. For each galaxy in a given
redshift and mass bin (i.e. $3.5\le z\le 4.5$ and $M_1\le M\le M_2$),
we summed up all the individual $PDF(M|z)$ in the given redshift
interval, after normalizing them to unit probability. This procedure
gives the probability $P(M_j,M_i)$ for a galaxy with an observed mass
$M_j$ (resulting from the best fit of its SED) to have a mass $M_i$
still compatible with its photometry and photometric redshifts
$PDF(z)$ derived from the Bayesian analysis, as described in Section
2.3.

\begin{figure}
\includegraphics[width=9cm,angle=0]{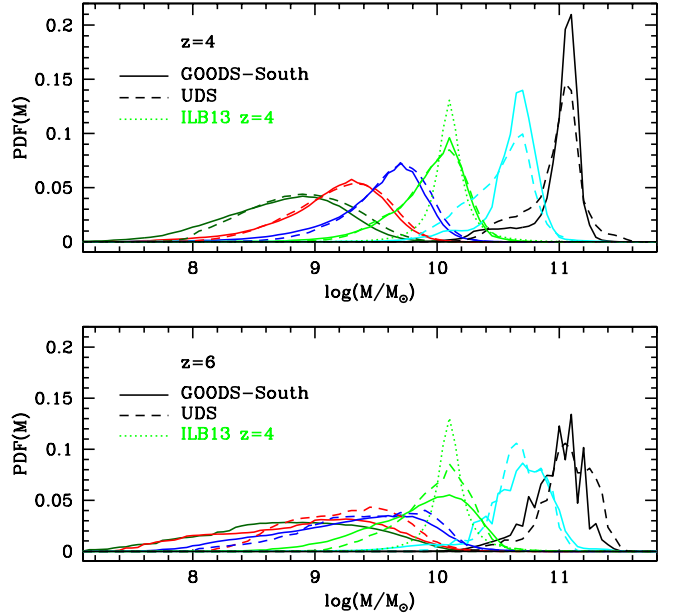}
\caption{
The probability distribution functions (PDFs) of stellar mass for
galaxies with measured stellar mass $M$ at different masses and
redshifts, resulting from the Monte Carlo simulations described in the
text. The PDFs were averaged over all galaxies contained in
contiguous bins with separation in mass of 0.2 in $log(M)$, although
we plot here only a few examples. The upper panel presents the PDFs at
$z=4\pm 0.5$, the lower panel at $z=6\pm 0.5$. In both panels the
dark-green curves are associated with galaxies in the bin centered at
$log(M/M_\odot)=8.9$, while red, blue, green, cyan and black curves
are associated to galaxies with masses $log(M/M_\odot)$ of 9.3, 9.7,
10.1, 10.7 and 11.1 respectively. Solid lines refer to
GOODS-South, dashed to UDS. The dotted green line shows for
comparison the PDF adopted at all masses by \cite{ilbert13} at $z=4$.
}
\label{sumpm}
\end{figure}

Fig.\ref{sumpm} shows the probability distribution functions for
the GOODS-South and UDS fields in different mass bins, both at $3.5\le
z\le 4.5$ (top) and at $5.5\le z\le 6.5$ (bottom).
The
dark-green curves are associated with galaxies in the bin centered at
$log(M/M_\odot)=8.9$, while red, blue, green, cyan and black curves
are associated to galaxies with masses $log(M/M_\odot)$ of 9.3, 9.7,
10.1, 10.7 and 11.1 respectively.
For comparison, we plot on the same
figure also the $PDF(M)$ adopted by \cite{ilbert13} at $z=4$, which is
the product of a Gaussian with $\sigma=0.5$ and a Lorentzian
distribution $L(x)=\frac{\tau}{2\pi}\frac{1}{(\tau/2)^2+x^2}$ with
$\tau=0.04*(1+z)$. The CANDELS PDFs shown in Fig.\ref{sumpm} are based on the
BC03 stellar library and on the choice of physical parameters (star
formation histories, age, metallicity, dust extinction, IMF) adopted
in this paper. The PDFs could vary slightly by adopting different
ingredients. However, as shown in the main text, the uncertainties due
to photometric scatter and photometric redshift uncertainties are much
larger than those due to different choices of the physical
ingredients.

From this plot we can draw some conclusions. First, and most
important, the error in the mass estimation is not constant at all
masses, as usually assumed (e.g. \cite{ilbert13}): it is indeed
smaller for higher-mass galaxies than for lower ones. This is expected
since larger photometric errors lead to wider ranges of acceptable
photometric redshifts and spectral models. We also note that the
error is not symmetric and, especially at $M\ge 10^{9-10}\, {\rm M_\odot}$
starts to show a tail towards lower masses. The second important
aspect is that at higher redshifts the combined PDFs are wider and the
asymmetry is more pronounced than at $z=4$, due to a combination of
larger photometric uncertainties and progressive dimming of the
galaxies for a given stellar mass. Thus, this plot shows the relative
enhancement of the uncertainties in PDFs towards low-mass objects,
independent of the adopted method for the stellar mass derivation.
We finally note that the $PDF(M|z)$ adopted by \cite{ilbert13} is
smaller than our own, at the same mass and redshift, and
substantially smaller than our own for faint galaxies. Given the
higher S/N and quality of our photometric data, this may likely reflect a
more conservative estimate of the implied errors in our computation, which
have been derived adopting a different technique w.r.t. \cite{ilbert13}.

Armed with this full characterization of the error on the estimated
mass, we can evaluate the impact of such errors on the estimate of the
(binned) GSMF. This is accomplished by convolving any input GSMF with
the error distribution of Fig.\ref{sumpm}, at the corresponding
redshift. For any given input GSMF we compute its expected values
$\Phi(M_j)$ in the same mass bins (with a step of 0.2 in $log(M)$)
used to derive the various $PDF(M|z)$. The expected mass function in
output is $\Phi_{conv}(i)=\Sigma_{j=1}^{N} \Phi(j)P(M_j,M_i)$, where
$N$ is the number of bins adopted to compute the $PDF(M|z)$.

\begin{figure}
\includegraphics[width=9cm,angle=0]{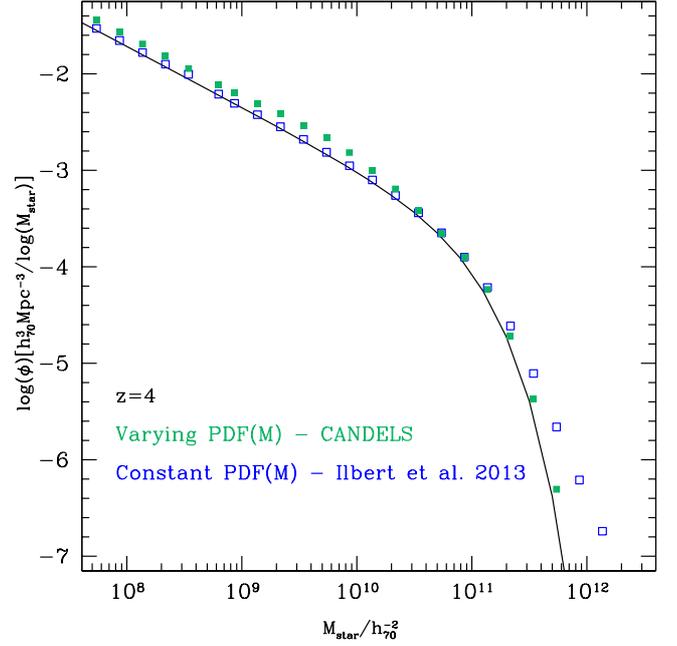}
\caption{
The effect of different prescription of the Eddington bias on the
observed GSMF. The black line shows a Schechter function representing
a GSMF with $\alpha\simeq -1.6$ and $M_*\simeq 11$. The open blue
squares show the resulting GSMF in bins of 0.2 in $logM$ after convolution
with a constant PDF at all masses, as adopted by \cite{ilbert13} at
$z=4$. The green filled squares represent the GSMF after convolution
with the more realistic mass-dependent PDF that widens when mass
decreases as we find in CANDELS (Fig.\ref{sumpm}).
}
\label{convphiz4ilb}
\end{figure}

To illustrate the effect of this procedure, and the differences with
respect to the previous analysis, we first take a representative input
mass function (with $\alpha\simeq -1.6$ and $M_*\simeq 11$) and
convolve it with the PDF of \cite{ilbert13} at $z=4$, namely a constant
function at all masses. Fig.\ref{convphiz4ilb} shows the convolved
mass function (empty blue squares) resulting from the intrinsic mass
function (solid line) after applying the convolution process described
above. We obtain a behaviour for the convolved GSMF similar to what
has been found by \cite{ilbert13} (their Fig.A.2), namely that the
low-mass side is unaffected by the Eddington bias while the density at
the high-mass end is enhanced.

Then, we consider an error distribution similar to that found in
CANDELS, i.e. with larger uncertainties at lower masses
(Fig.\ref{sumpm}). We adopt the same functional form of
\cite{ilbert13}, namely the product of a Gaussian and a Lorentzian
distributions, but we let the $\sigma$ of the gaussian change as a
function of mass in order to coarsely reproduce the observed PDFs at $z=4$
(hence being smaller than \cite{ilbert13}'s one at $log(M/{\rm M_{\odot}}>10.3$
and larger at smaller masses). Fig.\ref{convphiz4ilb} shows with filled green
squares the GSMF resulting from the same intrinsic mass function
(solid line) adopted in Fig.\ref{convphiz4ilb} after the convolution
process with a variable PDF in mass.

The combination of wider PDFs at lower masses and asymmetric
distributions has an interesting behaviour on the GSMF shape: in the
low-mass regime, $M\le 10^9 {\rm M_\odot}$, the error is symmetric and
contributes to a slight enhancement in the number density of galaxies
at the low-mass end, typically fitted with a power law. This results
into a steepening of the GSMF. According to \cite{eddington1913}, the
effect is higher for steeper distributions and for larger errors. At
the high-mass side, instead, the errors in mass are smaller and show
an asymmetry towards lower masses. As a consequence, the exponential
tail of the mass function is less affected by this scatter, and the
observed data points are a good representation of the intrinsic GSMF,
at least at $z=4$ for the CANDELS GOODS-South and UDS fields. This
behaviour is markedly different from what is derived assuming instead
a constant error on $PDF(M|z)$, even if the average error is adopted.

\begin{figure}
\includegraphics[width=9cm,angle=0]{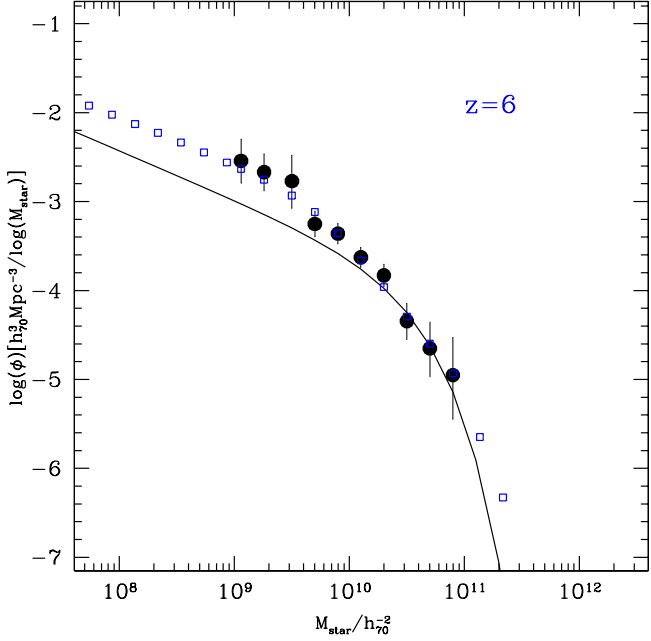}
\caption{
The effect of the Eddington bias on the observed GSMF at $z=6$ (black
dots). The black line shows a Schechter function representing the
resulting best fit GSMF at $z=6$. The open blue squares show the
resulting GSMF after convolution with the observed PDFs at $z=6$
(Fig.\ref{sumpm}, bottom panel).
}
\label{convphiz6}
\end{figure}

Moving to higher redshift, we find that at $z=5$ the situation is
similar to $z=4$, while at $z=6$ the effect of noise in the mass
estimate becomes more severe and hence the correction for the
Eddington bias is larger and more uncertain, as shown in
Fig.\ref{sumpm} (lower panel). Fig.\ref{convphiz6} shows the effect of the
Eddington bias correction at $z=6$. The wide uncertainties in mass, both
for faint and for bright objects, affect the GSMF at all scales, producing
a steepening of the low-mass side and a pronounced increase of the
exponential tail at high masses.

For instance, we find that there is a small but non-negligible
probability ($\sim 10^{-3}$) for a galaxy at $z=6$ to be scattered from
a mass of $10^{9.3} {\rm M_\odot}$ to $10^{10.5} {\rm M_\odot}$, while it is $\sim
10^{-6}$ at $z=4$.

At $z\simeq 7$ these effects become so large that a proper treatment
of the Eddington bias is simply impossible with the present data. For
instance, the probability that the same galaxy at a mass of $10^{9.3}
{\rm M_\odot}$ is scattered to $10^{10.5} {\rm M_\odot}$ is as large as $\sim
10^{-2}$. Thus if we observe a density of galaxies of $\Phi=10^{-5}
{\rm Mpc^{-3}Mag^{-1}}$ at $M=10^{10.5} {\rm M_\odot}$, this can be entirely due
to galaxies at $M=10^{9.3} {\rm M_\odot}$ (with $\Phi=10^{-3}
{\rm Mpc^{-3}Mag^{-1}}$) that are scattered to higher masses due to
uncertainties in their stellar mass estimation. Since this correction
is so important at the high-mass end of the $z>6$ GSMF and the
derivation of the PDFs at such low levels of probability depends
critically on the details (photometric redshifts, stellar libraries
adopted, star formation histories, grid of age, dust, metallicity), we
can conclude that at the present stage the Eddington bias correction
at $z=7$ is highly uncertain.
Moreover, at $z=7$ the PDFs are very noisy due also to the low number
statistics (at mass greater than $10^{10.3} {\rm M_\odot}$ we have only 4 galaxies
in the whole GOODS-South and UDS fields).

The derivation of the best-fitting Schechter functions have been
carried out using the formalism described above. For any possible
combination of the Schechter parameters $\alpha$, $M^*$, and $\Phi^*$,
we compute the convolved GSMF using the observed PDFs in mass and we
compare it with the observed mass function. We scan the three
parameters of the Schechter function to find the best fit solution by
a $\chi^2$ minimization. The GSMFs presented in the main text have
been computed accordingly.

We note that, for the reasons described above, we decided not to apply
the proper correction for the Eddington bias in the $z=7$ GSMF. We
adopt at $z=7$ the same PDFs derived at $z=6$, which are less noisy,
as a conservative assumption. The small range in masses sampled by our
GSMF at $z=7$ results in large uncertainties in the best fit Schechter
function parameters due to degeneracies between $\alpha$, $M^*$, and
$\Phi^*$, as shown in Fig.\ref{alpha}. We have verified that due to
these degeneracies the parameter space allowed at 1$\sigma$ by the
present data is wide and it does not depend strongly on whether we
adopt the PDFs at $z=6$ or the ones determined at $z=7$ to correct the
Eddington bias in the redshift range $6.5<z<7.5$.

\section{The impact of neglecting to correct for Eddington bias at
high redshift}

In Section 6 we have shown that the uncertainties in the
measurement of the galaxy stellar mass do produce a systematic effect
in the output GSMF, that must be taken into account when one derives
the best-fitting Schechter parameters. However, we have also shown
that accurate corrections for this effect are difficult to estimate,
partly because they are model-dependent and partly because of the
limited statistics available, especially at the highest redshifts.

For the sake of completeness, we report here the results on the GSMF
fitting without the corrections for the Eddington bias. This is
useful to understand what are the uncertainties at work when one is
dealing with the mass function, and to warn the reader about the
consequences of neglecting or underestimating this correction. We
would like to stress that, given the very high quality of the data
used here, other surveys with data of lower S/N or narrower wavelength
range are even more affected.

\begin{figure}
\centering
\includegraphics[width=7cm,angle=-90]{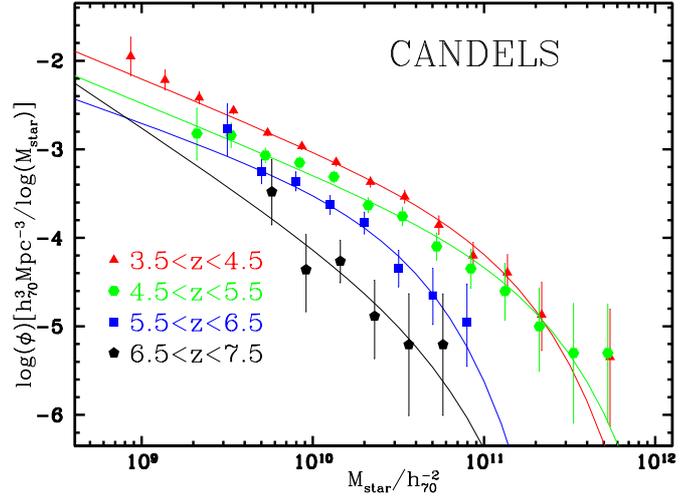}
\caption{
The GSMFs from $z=4$ to $z=7$ in the CANDELS UDS and GOODS-South
fields. At variance with the main paper, we have neglected here the
effects of the uncertainties in the stellar mass (the so--called
Eddington bias). The error bars take into account the Poissonian
statistics and the uncertainties derived through the Monte Carlo
simulations. The solid continuous curves show the best-fitting
Schechter function.
}
\label{summarymfhzapp}
\end{figure}

\begin{figure}
\includegraphics[width=9cm,angle=0]{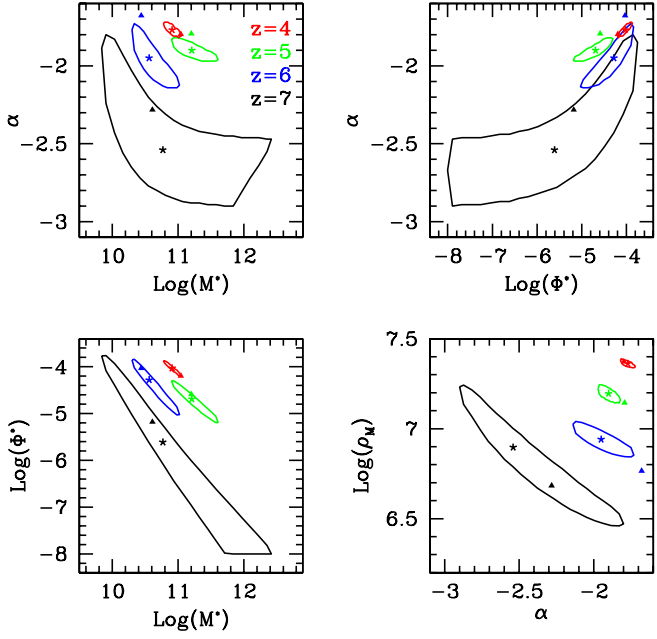}
\caption{The evolution of the three parameters ($\alpha$, $M^*$, $\Phi^*$)
of the GSMF with redshift, neglecting the effect of the Eddington bias.
The bottom-right panel shows the dependencies of
the Stellar Mass Density ($\rho_M$ in unit of $M_{\odot}Mpc^{-3}$) from
the parameter $\alpha$. The stars mark the position of the best fit
of the observed GSMF with a Schechter function, while the triangles indicate
the position of the best fit from the Maximum Likelihood procedure.
}
\label{alphaapp}
\end{figure}

Following the technique described in the previous sections, but
removing any correction for the Eddington bias, we have obtained the
GSMF that is shown in Fig.\ref{summarymfhzapp}. The derived
uncertainties on the Schechter function parameters $\alpha$, $M^*$ and
$\Phi^*$, are also shown in Fig.\ref{alphaapp}.

\begin{table}
\caption{Mass Function best fit parameters}
\label{tabgsmfapp}
{\centering
\begin{tabular}{c | c c c r}
\hline\hline
Redshift & $\alpha$ & $\log(M^*)$ & $\log(\Phi^*)$ & $N_{gal}$ \\
\hline
$3.5<z<4.5$ & $-$1.77$\pm$0.05 & 10.91$\pm$0.14 & $-$4.03$\pm$0.17 & 1293 \\
$4.5<z<5.5$ & $-$1.90$\pm$0.08 & 11.21$\pm$0.36 & $-$4.69$\pm$0.44 &  370 \\
$5.5<z<6.5$ & $-$1.95$\pm$0.20 & 10.56$\pm$0.36 & $-$4.28$\pm$0.60 &  126 \\
$6.5<z<7.5$ & $-$2.54$\pm$0.55 & 10.77$\pm$1.29 & $-$5.61$\pm$2.13 &   20 \\
\hline
\end{tabular}
}
\\
The best-fit parameters of the Schechter function that has been fitted
to the observed GSMF, when the effects of uncertainties in the stellar
mass (the so--called Eddington bias) are neglected.
\end{table}

It is immediately clear that the results are significantly different
from our main analysis. The slope $\alpha$ is steeper than our best
fit and further steepens with redshift moving from $\alpha=-1.8$ at
$z=4$ to $\alpha \simeq -2$ at $z=6$. The characteristic mass
$\log(M^*)$, on the contrary, evolves only marginally from $z=4$ to
$z=6$. This is exactly what is predicted by our analysis of the Eddington
bias, that is expected to artificially steepen the slope (at all
redshifts) and progressively increase the GSMF in our higher redshift
bins.

\begin{figure}
\includegraphics[width=9cm,angle=0]{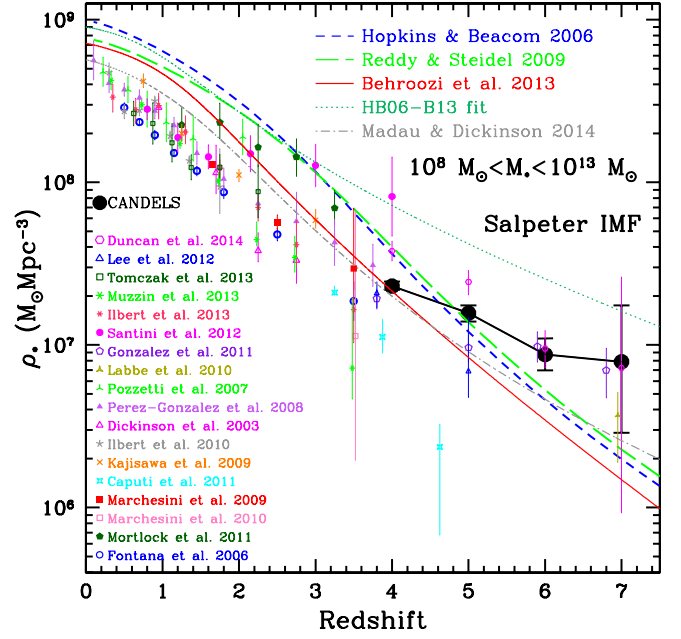}
\caption{
The redshift evolution of the stellar mass density (SMD) at
$3.5<z<7.5$ obtained in the CANDELS UDS and GOODS-South fields
presented in this paper (black points), when the effect of
uncertainties in the stellar mass (the so-called Eddington bias) are
not taken into account. The SMD is compared to the lower redshift
data from different surveys. $\rho_M$ is in units of ${\rm
M_{\odot}Mpc^{-3}}$ and has been obtained by integrating the best
fit mass functions from $M_{min}=10^8 {\rm M_{\odot}}$ to
$M_{max}=10^{13} {\rm M_{\odot}}$.
All the SMDs have been converted to a Salpeter IMF for comparison.
The error bars of the CANDELS data have been computed
using the same Monte Carlo simulations developed to derive the
uncertainties on the Schechter function parameters.  The short-dashed
line is the stellar mass density obtained integrating over cosmic time
the star formation rate density (SFRD) of \cite{hb06}.  The
long-dashed line is the SMD from the SFRD of \cite{rs09}. The solid
line is the SMD obtained by integrating the SFRD of \cite{behroozi13},
while the dotted line is the SMD derived by integrating the new fit of
the \cite{hb06} carried out by \cite{behroozi13}. The dotted-dashed
line shows the SMD derived from the SFRD given by of \cite{md14}.  All
the stellar mass densities obtained by integrating the different SFRDs
assume a constant recycling fraction of 28\%.
}
\label{smdapp}
\end{figure}

These changes are also reflected in the derived evolution of the
stellar mass density that is reported in Fig.\ref{smdapp}. As a result
of the steeper slope and higher $M^*$, the resulting $\rho_M$ is
significantly above the integrated evolution of the Star--Formation
Rate Density.

These results underline the importance of a careful description of the
Eddington bias in the estimate of the GSMF.

\begin{table}
\caption{Stellar Mass Density at $3.5<z<7.5$}
\label{tabsmdapp}
\centering
\begin{tabular}{c | c c c}
\hline\hline
Redshift & $log(\rho_M)$ & Min $log(\rho_M)$ & Max $log(\rho_M)$ \\
\hline
$3.5<z<4.5$ & 7.36 & 7.33 & 7.39 \\
$4.5<z<5.5$ & 7.20 & 7.14 & 7.24 \\
$5.5<z<6.5$ & 6.94 & 6.84 & 7.04 \\
$6.5<z<7.5$ & 6.90 & 6.46 & 7.24 \\
\hline
\end{tabular}
\\
The stellar mass density $log(\rho_M)$ is derived from the best fit of
the GSMF, neglecting the Eddington bias, and integrating it from
$M=10^{8} {\rm M_{\odot}}$ to $M=10^{13} {\rm M_{\odot}}$. A Salpeter
IMF is assumed.  The SMD $\rho_M$ is in units of ${\rm
M_{\odot}Mpc^{-3}}$.  The minimum and maximum SMDs indicate the
1-$\sigma$ range (i.e. the 68\% confidence interval).
\end{table}

\end{appendix}

\end{document}